\documentclass[aps,prb,twocolumn,showpacs,10pt,superscriptaddress]{revtex4-1}
\usepackage{amssymb}
\usepackage{graphicx}
\usepackage{amsfonts}
\usepackage{amsmath}
\usepackage[usenames]{color}

\begin{document}

\title{Theory of box-model hyperfine couplings and transport signatures of long-range nuclear-spin coherence in a quantum-dot spin valve}
\author{Stefano Chesi}
\affiliation{Beijing Computational Science Research Center, Beijing 100084, China}
\affiliation{Department of Physics, McGill University, Montr{\'e}al, Qu{\'e}bec H3A 2T8, Canada}
\author{W.~A.~Coish}
\affiliation{Department of Physics, McGill University, Montr{\'e}al, Qu{\'e}bec H3A 2T8, Canada}
\affiliation{Quantum Information Science Program, Canadian Institute for Advanced Research, Toronto, Ontario M5G 1Z8, Canada}

\date{\today}

\begin{abstract}
We have theoretically analyzed coherent nuclear-spin dynamics induced by electron transport through a quantum-dot spin valve. The hyperfine interaction between electron and nuclear spins in a quantum dot allows for the transfer of angular momentum from spin-polarized electrons injected from ferromagnetic or half-metal leads to the nuclear spin system under a finite voltage bias. Accounting for a local nuclear-spin dephasing process prevents the system from becoming stuck in collective dark states, allowing a large nuclear polarization to be built up in the long-time limit. After reaching a steady state, reversing the voltage bias induces a transient current response as the nuclear polarization is reversed.  Long-range nuclear-spin coherence leads to a strong enhancement of spin-flip transition rates (by an amount proportional to the number of nuclear spins) and is revealed by an intense current burst, analogous to superradiant light emission. The crossover to a regime with incoherent spin flips occurs on a relatively long time scale, on the order of the single-nuclear-spin dephasing time, which can be much longer than the time scale for the superradiant current burst. This conclusion is confirmed through a general master equation. For the two limiting regimes (coherent/incoherent spin flips) the general master equation recovers our simpler treatment based on rate equations, but is also applicable at intermediate dephasing. Throughout this work we assume uniform hyperfine couplings, which yield the strongest coherent enhancement. We propose realistic strategies, based on isotopic modulation and wavefunction engineering in core-shell nanowires, to realize this analytically solvable ``box-model'' of hyperfine couplings.

\end{abstract}

\pacs{68.65.Hb,72.25.Pn,75.78.-n}



\maketitle

\section{Introduction}

Continuous efforts toward the implementation of quantum computing with electron-spin qubits have led to several advances in the spin manipulation of single electrons in III-V semiconductor quantum dots and to improved understanding of the associated decoherence mechanisms.\cite{Loss1998,Hanson2007,Coish2009,Zak2010} From these studies, it has become clear that hyperfine interactions with nuclear spins in the host material typically limit electron spin coherence. Several approaches have been developed to limit this source of decoherence (including, e.g., nuclear-spin state narrowing through passive measurement\cite{Coish2004,Klauser2006,Stepanenko2006,Giedke2006,Shulman2014} or active feedback control\cite{Greilich2007,Latta2009,Vink2009,Xu2009,Bluhm2010a} of the nuclear Overhauser field, or through spin-echo and more general dynamical-decoupling techniques\cite{Petta2005,Koppens2008,Bluhm2010}). These approaches are often limited to an inconvenient parameter regime (e.g., large magnetic field) or require additional pulsed or continuous-wave excitation. For these reasons, new strategies to accurately control coupled electron-nuclear spin dynamics are being actively pursued. 

Quantum dots confining single holes have recently emerged as a promising alternative platform for spin qubits, since their Ising-like hyperfine coupling allows for superior control of coherence, relative to electron systems.\cite{Fischer2008,Eble2009,Gerardot2008,Brunner2009,Degreve2011,Wang2012,Chesi2014,Carter2014,Wang2014} Another strategy is to exploit group-IV materials: C, Ge, and Si, all of which can be isotopically enriched to be nuclear-spin free.\cite{Morton2011,Zwanenburg2012} However, the nuclear-spin bath can also serve as a useful resource, providing a highly local tunable effective magnetic field, or long-lived quantum memory, as demonstrated by the transfer of the qubit state from the electron to the nuclear-spin system in NV centers and phosphorus donors in Si.\cite{Sar2012,Pla2012,Pla2013} For quantum dots, being able to engineer a well-understood form of hyperfine coupling which allows better control of the coupled electron-nuclear spin dynamics could lead to improved quantum memories that exploit collective nuclear-spin degrees of freedom\cite{Taylor2003} and a fundamentally improved level of control over electron-spin coherence.

A major difficulty in modeling spin dynamics arises from the typical non-uniformity of the electron-nuclear hyperfine coupling strength. In fact, while many exact and approximate theoretical approaches to strongly-coupled electron-nuclear spin dynamics have been developed for various specific limits,\cite{Khaetskii2002, Khaetskii2003, Dobrovitski2003, Coish2004, Bortz2007, Cywinski2009, Coish2010, Rudner2010, Barnes2012, Wang2012, Faribault2012, Gullans2013} a controlled theory applicable to a typical number of $10^5$-$10^6$ nuclear spins does not exist for many experimentally relevant regimes (e.g., very low magnetic field). On the other hand, a simple exact solution based on total angular-momentum eigenstates can be found if the hyperfine coupling-constants are uniform, the so-called ``box-model''.\cite{Khaetskii2003,Coish2007,Zhang2006} Due to the simplicity of this limit, there is a large relevant body of theoretical literature, addressing electron-spin dephasing,\cite{Coish2007,Erbe2010,Barnes2011} as well as manipulation and entanglement generation/preservation for the electron-spin\cite{Christ2008,Mazurek2014,MazurekEPL2014,Bragar2014} or nuclear-spin system.\cite{Eto2002,Eto2004,Erbe2011,Rudner2011} 

The inhomogeneous hyperfine coupling in current devices is due to the spatial dependence of the electronic wavefunction. For a bound state, this inhomogeneity is thus impossible to avoid in III-V materials (where all isotopes have a finite nuclear spin). Although uniform coupling is often assumed with the purpose of gaining insight into realistic setups, it is not always clear to what extent these results are applicable to experimentally realizable situations. With these motivations in mind, here we discuss how Si/Ge core/shell nanowires can offer a promising route to realize the uniform limit of the hyperfine interaction. In the following sections, we describe quantum-dot designs based on this type of nanowire, which approach the uniform-hyperfine-coupling regime in a systematic fashion.

An immediate consequence of uniform ``box-model'' hyperfine couplings would be a strong enhancement of the spin-flip electric transport through quantum dots, induced by the flip-flop component of the hyperfine interaction. This phenomenon has been predicted in earlier works by Eto et al. \cite{Eto2002,Eto2004} and is analogous to superradiant photon emission of atoms with uniform coupling to the optical field.\cite{Dicke1954,Gross1982} An alternative setup with unpolarized contacts has also been recently analyzed in Ref.~\onlinecite{Schuetz2012}. These ideas are also relevant to quantum dots with magnetic impurities\cite{Besombes2004,Rossier2007,Lai2015} and, in addition to transport, optical superradiance due to the nuclear-spin system has also been proposed for single-photon emitters.\cite{Kessler2010} 

To maximize the influence of spin-flip contributions to current, here we will consider a quantum-dot spin valve with antiparallel ferromagnetic leads. Experimental realizations of quantum-dot spin valves include InAs quantum dots with Ni ferromagnetic contacts,\cite{Hamaya2007,Hamaya2008a,Hamaya2008b} as well as a spin valve based on a Si nanowire,\cite{Tarun2011} which is compatible with our proposed implementation of uniform hyperfine interaction. There is also ongoing interest in carbon nanotubes with ferromagnetic contacts, including quantum dot spin valves (see, e.g., the recent Ref.~\onlinecite{Stadler2015,Dirnaichner2015} and references therein). With this experimental progress in mind, here we examine a scheme to demonstrate superradiance-like behavior in a quantum-dot spin valve with ferromagnetic leads. To account for imperfectly-polarized ferromagnetic contacts, we allow for nuclear-spin states that are both partially polarized and fully dephased, as well as spin-flip tunneling processes allowing for spin flips in both directions. These aspects are not present in the superradiant-like transport in an ideal `spin-blocked' regime discussed in previous literature\cite{Eto2002,Eto2004,Schuetz2012} (in our case, for half-metal leads).  These features are, however, generically relevant for most realistic setups. Furthermore, the quantum-dot spin valve discussed here offers certain advantages with respect to control over the nuclear spin polarization. In particular, in the setup described here, it is possible to reverse the nuclear-spin magnetization direction by reversing the bias. This allows for an efficient initialization mechanism (without inverting the magnetic field), which is required to detect a strong enhancement of the transient current to provide evidence of long-range nuclear-spin coherence. 

This paper is organized as follows: in Sec.~\ref{sec:hyperfine} we define the hyperfine interaction with uniform coupling strength and in Sec.~\ref{sec:SiGenanowires} we describe possible strategies to approach this limit, based on Si/Ge core/shell nanowire quantum dots. The rest of the paper analyzes a quantum-dot spin valve setup described in Sec.~\ref{sec:spinvalve}. We introduce relevant tunneling rates in Sec.~\ref{sec_rates} and discuss the superradiance analogy for half-metal leads in Sec.~\ref{sec_half_metal}. In Sec.~\ref{sec_nuclear_evolution} we characterize the spin valve in the two extreme limits of nuclear-spin dephasing (fully coherent/incoherent nuclear-spin evolution). We analyze stationary states, magnetization dynamics, and the electron current. In Sec.~\ref{sec_mastereq} the two limiting regimes of Sec.~\ref{sec_nuclear_evolution} are obtained from a more general master equation, which also allows us to discuss the case of intermediate dephasing. Section~\ref{sec_concl} contains our final remarks and in Appendices~\ref{sec:exact_eigenstates} and \ref{sec:appendix} we provide some additional details on the exact eigenstates with uniform hyperfine coupling and on the master equation approach of Sec.~\ref{sec_mastereq}.

\section{Hyperfine interaction}\label{sec:hyperfine}

The Fermi contact interaction of a single electron with its nuclear bath is well known:
\begin{equation}\label{hyperfine_inhomo}
H_{\rm hf} = \sum_k A_k {\bf I}_k \cdot {\bf S},
\end{equation}
where $\mathbf{S}=\frac{1}{2}\sum_{\sigma,\sigma^{\prime}}\boldsymbol{\sigma}_{\sigma \sigma^{\prime}}d_{\sigma}^{\dagger} d_{\sigma^{\prime}}$ is the electron spin operator (we set $\hbar=1$), $\boldsymbol{\sigma}$ is the vector of Pauli matrices, $d^\dag_\sigma$ creates an electron with spin $\sigma=\uparrow,\downarrow$ in the lowest orbital of a quantum dot, and ${\bf I}_k$ is the spin operator for nucleus $k$. The hyperfine coupling is $A_k= v_0 A^{j_k}|\psi({\bf r}_k)|^2$, where $j_k$ indicates the isotopic species, $v_0$ is the atomic volume, and $\psi({\bf r}_k)$ is the value of the electronic envelope wave function at the nuclear site ${\bf r}_k$. 

The central spin problem resulting from Eq.~(\ref{hyperfine_inhomo}) is in general very complex to analyze.\cite{Khaetskii2002, Khaetskii2003, Dobrovitski2003, Coish2004, Bortz2007, Cywinski2009, Coish2010, Barnes2012, Wang2012, Faribault2012,Zak2010} The limit of $N$ nuclei with uniform hyperfine couplings (which allows for an exact solution) is therefore especially interesting:
\begin{align}
H_{\rm hf} & = H^{\rm hf}_{zz} + H^{\rm hf}_{ff},\\
H^{\rm hf}_{zz} & =\frac{A}{N}\, I_z S_z,  \quad H^{\rm hf}_{ff}= \frac{A}{2N} (I_+ S_- + I_- S_+),\label{hf}
\end{align}
where ${\bf I}=\sum_k {\bf I}_k$ is the total nuclear angular momentum, $I_\pm=I_x\pm i I_y$, $S_\pm=S_x\pm i S_y$, and, for later convenience, in Eq.~(\ref{hf}) we have explicitly written  the secular and flip-flop contribution, $H^{\rm hf}_{zz}$ and $H^{\rm hf}_{ff}$, respectively. An exact solution for the dynamics of $H_{\rm hf}$ can be found in terms of total nuclear angular-momentum eigenstates $|I, I_z\rangle$ (see, e.g., Refs.~\onlinecite{Eto2002,Khaetskii2003,Eto2004} for a solution in the absence of a magnetic field and Refs.~\onlinecite{Coish2007,Zhang2006} for the case with a magnetic field), where we omit here and in the following an additional permutation quantum number.\cite{Arecchi1972}   The underlying assumption of Eq.~(\ref{hf}) is
\begin{equation}\label{uniform_A}
A^{j_k}|\psi({\bf r}_k)|^2 = \left\{
\begin{array}{ll}
\frac{A}{v_0 N} & {\rm for} ~1\leq k \leq N\\
0 & {\rm otherwise}
\end{array} \right. .
\end{equation}
This ``box-model'' is difficult to implement in actual devices. For the most-studied example of GaAs quantum dots, Eq.~(\ref{uniform_A}) becomes impossible to realize, since $A^{j} \neq 0$ for all isotopes of Ga and As and a uniform wavefunction $\psi({\bf r})$ which abruptly vanishes outside the quantum dot is not realistic. On the other hand, it is possible to devise strategies approaching Eq.~(\ref{uniform_A}) in Si/Ge core/shell structures, a promising platform for quantum information processing.\cite{Hu2007,Hu2011,Kloeffel2011}

\section{Uniform hyperfine coupling in Si/Ge core/shell nanowires}\label{sec:SiGenanowires}

\begin{figure}
\includegraphics[width=0.47\textwidth]{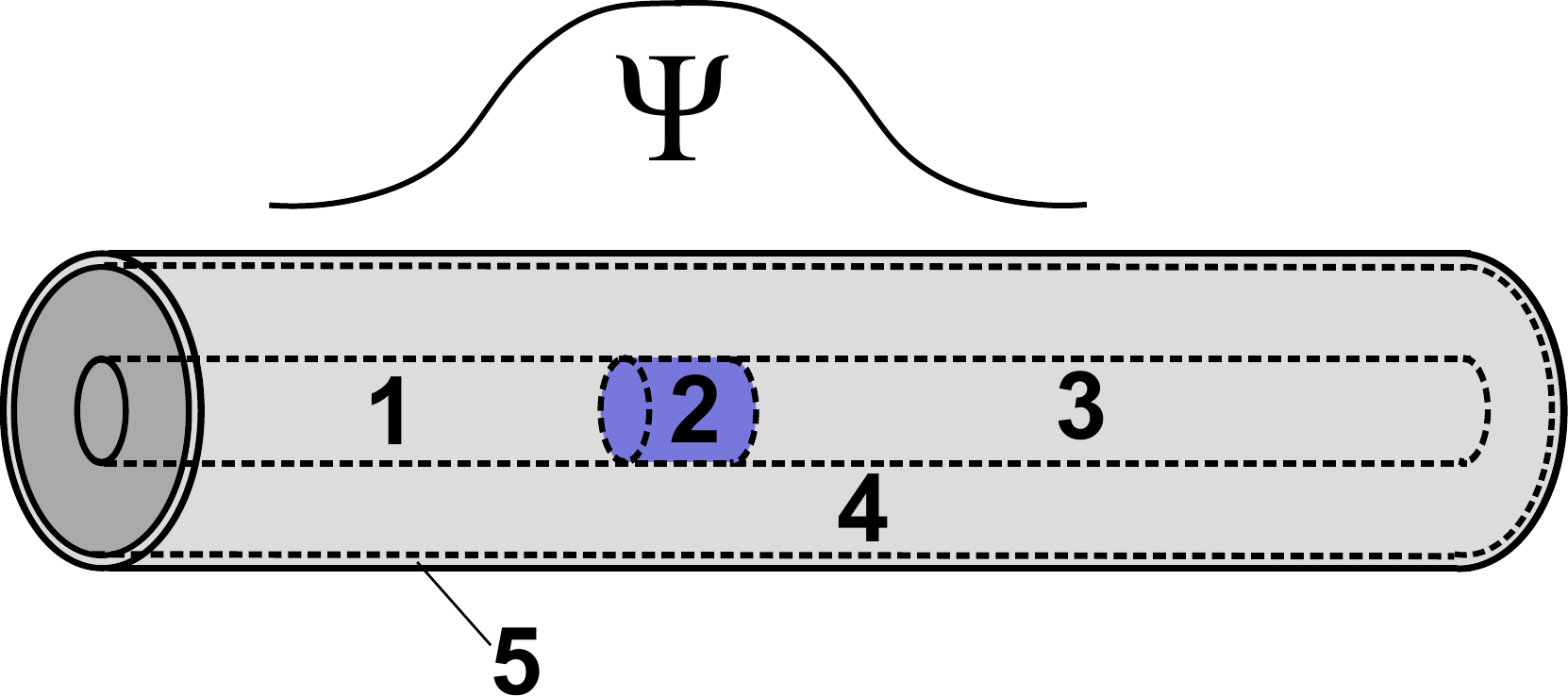}
\caption{\label{nanowire} Schematic of a nuclear-spin-free core/shell structure with an embedded island of nuclear spins. The core is composed of regions 1-2-3, where the middle section (blue region 2) is enriched by a spinful isotope. Regions 4 and 5 represent different shells of the nanowire. By using additional gates along the nanowire the electron wavefunction $\psi({\bf r})$ is centered around region 2, such that the  hyperfine couplings are approximately uniform.}
\end{figure}

To realize a uniform hyperfine coupling, it is necessary that the host crystal has both isotopes with and without nuclear spin, a condition which is indeed satisfied for both Ge and Si. We have found that core/shell nanowires offer the possibility to approach the ideal situation of uniform hyperfine couplings, see Eq.~(\ref{uniform_A}). The general structure of such a nanowire is illustrated in Fig.~\ref{nanowire}, with a core and multiple shells of different SiGe alloys in the radial direction.\cite{Lauhon2002,Goldthorpe2009,Hu2012}

\subsection{Embedded islands of nuclear spins}

A possible design achieving almost uniform hyperfine interaction is illustrated in Fig.~\ref{nanowire} and is based on including nuclear spins only in a small region of the wire (region 2). The structure is realistic to grow, being analogous to embedded quantum dots,\cite{Borgstrom2005,Kats2012} where the embedded island is of a different chemical composition than the surroundings. Instead, the same alloy can be used here for all regions 1-4 (with shell 5 providing confinement), but region 2 is isotopically enriched with nuclear spins. Isotopic modulation of silicon nanowires has already been demonstrated in Ref.~\onlinecite{Moutanabbir2011}.

Additional electrostatic gates make it possible to localize a single electron with wavefunction centered at the island of nuclear spins, see Fig.~\ref{nanowire}. Although this does not strictly implement Eq.~(\ref{uniform_A}), the hyperfine coupling is approximately uniform for a wavefunction $\psi({\bf r})$ of comparatively large extent with respect to the size of the active central region. A typical core diameter is 10 nm and a few atomic layers along the nanowire, with several hundred nuclear spins each, can be strongly coupled to a bound electron spin. 

In addition to the nearly uniform coupling, a further advantage of having an isolated island of nuclear spins is that diffusion of nuclear spin polarization out of the quantum dot becomes impossible (in contrast, e.g., to GaAs quantum dots). Finally, the electron can be moved away from the region 2 through electrostatic gates, turning off the contact hyperfine interaction.

\subsection{Wavefunction engineering}

The strategy indicated in the previous section relies on confining the nuclear spins to a small region in which $\psi({\bf r})$ is approximately uniform, but now we describe a possible route to realize wavefunctions with large uniform regions. In principle, the desired $\psi({\bf r})$ can always be defined through suitable confinement, as can be seen for one-dimensional systems. Given a ground-state wavefunction $\psi(z)$ with energy $E_0$, the corresponding potential is:
\begin{equation}\label{potential_from_psi}
V(z)=\frac{1}{2m\psi(z)}\frac{\partial^2 \psi(z)}{\partial z^2}+E_0,
\end{equation} 
with $V(z)=E_0$ constant in the central region where $\psi(z)$ is uniform. A specific example is simply obtained from the following $\psi(z)$, defined for $-W<z<W$ and uniform in the central region $-W+\frac{d}{2} < z < W-\frac{d}{2}$:
\begin{equation}\label{psi_z_uniform}
\psi(z) \propto 
\left\{
\begin{array}{cl}
\sin \pi(z+W)/d & {\rm for}~ -W < z \leq -W+\frac{d}{2} ,\\
1~~& {\rm for}~ -W+\frac{d}{2} < z < W-\frac{d}{2} ,\\
\sin \pi(W-z)/d & {\rm for}~ W-\frac{d}{2} \leq z < W .
\end{array}
\right.
\end{equation} 
This wavefunction corresponds to the following potential:
\begin{equation}\label{V_z_uniform}
V(z) =
\left\{
\begin{array}{cl}
-\frac{\pi^2}{2md^2} & {\rm for}~ -W < z \leq -W+\frac{d}{2} ,\\
0~~& {\rm for}~ -W+\frac{d}{2} < z < W-\frac{d}{2} ,\\
-\frac{\pi^2}{2md^2} & {\rm for}~ W-\frac{d}{2} \leq z < W,
\end{array}
\right.
\end{equation} 
where we have have set $E_0=0$ in Eq.~(\ref{potential_from_psi}). With a different choice of ground-state wavefunction, it is also easy to obtain from Eq.~(\ref{potential_from_psi}) a smooth potential $V(z)$ analogous to Eq.~(\ref{V_z_uniform}), i.e., with a central region in which $V(z)=0$ and two negative potential wells on both sides of such a central region.

A similar strategy can be applied to the radial confinement and, in fact, core/shell structures allow for the realization of a stepwise potential analogous to Eq.~(\ref{V_z_uniform}). The basic idea is to take advantage of negative band offsets and is illustrated in Fig.~\ref{nanowire}, where we assume now that the core (regions 1, 2, 3) and the inner shell (region 4) each has a distinct chemical composition (e.g., two different SiGe alloys) such that the band offset $-V_0$ of the inner shell (region 4) is negative with respect to the core. We can estimate a typical thickness of the shell by using band parameters of electrons in Ge and by assuming for simplicity that both the nanowire growth direction and band minimum (valley) are along [111], such that the transverse mass is $m_\perp \simeq 0.08 m_0$ (with $m_0$ the free electron mass; note that the valley degeneracy is generally broken by the presence of strain and confinement). In this case, the radial motion is described by:
\begin{equation}\label{H_radial}
H_\perp= \frac{{\bf p}_\perp^2}{2 m_\perp} -V_0 \theta(r-R_{\rm in})\theta(R_{\rm out}-r)+U(r),
\end{equation}
where $R_{\rm in/out}$ is the inner/outer radius of region 4 and typical values of $V_0$ for SiGe compounds can reach up to several hundred meV.\cite{VdWalle1986,Lu2005} If the outer shell (region 5) has a large positive band offset, it can be approximated as an infinite barrier
\begin{equation}
U(r)=\left\{ 
\begin{array}{cl}
0 & {\rm for}~ r< R_{\rm out}, \\
\infty & {\rm for}~ r \geq R_{\rm out} .\\
\end{array}
 \right. 
\end{equation}
%
\begin{figure}
\includegraphics[width=0.47\textwidth]{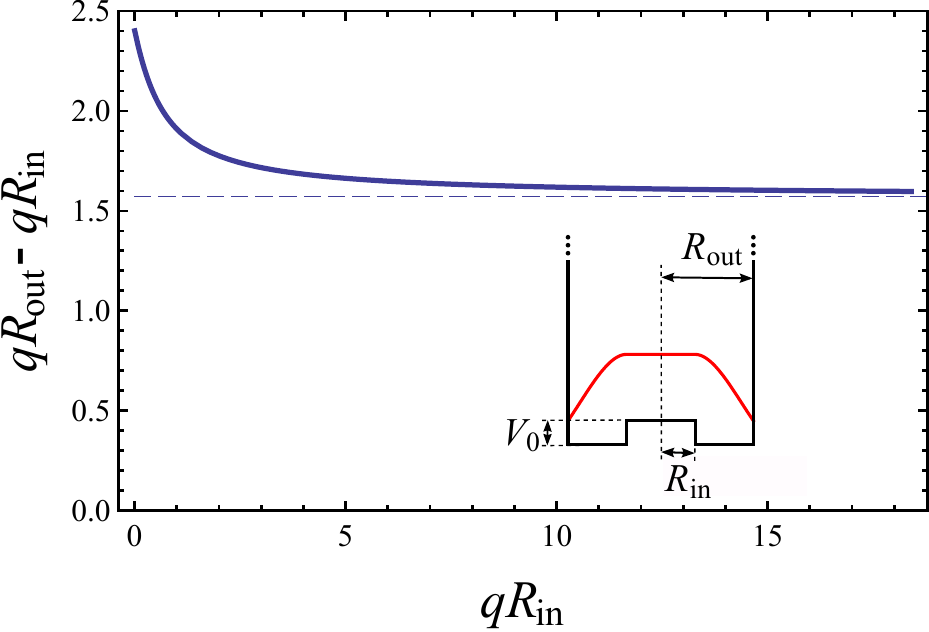}
\caption{\label{shell_radius} Value of $qR_{\rm out}-qR_{\rm in}$ as a function of $q R_{\rm in}$, as given by Eq.~(\ref{radius_out}). The dashed line is the asymptotic value $\pi/2$, see Eq.~(\ref{radius_approx}). The inset illustrates the radial dependence of the confining potential and ground-state wavefunction. The length scale for $R_{\rm in/out}$ is given by $q^{-1}=\hbar/\sqrt{2m_\perp V_0}\simeq 5$ nm, assuming a band offset $V_0=20$ meV and $m_\perp=0.8 m_0$. With these parameters, the asymptotic value of the shell thickness (corresponding to the dashed line) is $R_{\rm out}-R_{\rm in}\simeq 7.7$ nm. }
\end{figure}
%
The profile of the total radial potential is schematically represented in the inset of Fig.~\ref{shell_radius}. As in the one-dimensional problem, the requirement of a uniform wavefunction in the core corresponds to a zero-energy ground-state. Thus, the ground-state wavefunction has the form $\alpha J_0(qr)+ \beta Y_0(qr)$ in region 4, where $J_n(x)$ and $Y_n(x)$ are Bessel functions of the first and second kind, respectively, and $q=\sqrt{2m_\perp V_0/\hbar^2}$. Imposing the relevant boundary conditions at both $R_{\rm in}$ and $R_{\rm out}$ gives
\begin{equation}\label{radius_out}
J_0(q R_{\rm out}) Y_1(q R_{\rm in})- Y_0(q R_{\rm out}) J_1(q R_{\rm in})=0.
\end{equation}
For a given core radius $R_{\rm in}$ and chemical composition (i.e., the offset $V_0$), Eq.~(\ref{radius_out}) determines the appropriate thickness $R_{\rm out}-R_{\rm in}$. The result is plotted in Fig.~\ref{shell_radius} as a function of $q R_{\rm in}$. At $q R_{\rm in}=0$ the value $q R_{\rm out} \simeq 2.405$ is obtained from the first zero of $J_0(x)$. At large values of $q R_{\rm in}$ the asymptotic result is
\begin{equation}\label{radius_approx}
qR_{\rm out}-qR_{\rm in} \simeq \frac{\pi}{2}.
\end{equation}
As a numerical example, $V_0 =20$ meV gives $q^{-1} \simeq 5$ nm (the length scale of Fig.~\ref{shell_radius}). For this value, $R_{\rm in} = 10$ nm corresponds to a shell thickness of $8.6\,\mathrm{nm}$.  In this example, the shell thickness will approach the value $7.7\,\mathrm{nm}$, from Eq.~(\ref{radius_approx}), for larger values of $R_{\rm in}$. 

Core/shell structures can have a variety of designs and there should be no fundamental limitation to reach a high level of accuracy in fabrication (comparable to III-V planar heterostructures). We will therefore not pursue a more specific analysis of an ideal setup, but will simply assume in the following that a strategy similar to that given here should allow for an accurate realization of Eq.~(\ref{hf}). We discuss next how the uniformity of the hyperfine coupling can be revealed through electric transport measurements.

\section{Spin valve setup}\label{sec:spinvalve}

A quantum-dot spin valve consists of a quantum dot in contact with two ferromagnetic reservoirs with opposite polarization. The isolated dot and leads can be described through the Hamiltonian
\begin{equation}\label{H_ideal}
H_{\rm el} = 
\sum_\sigma \Big(V_g -\frac{\sigma b}{2}\Big) d_\sigma^\dag d_\sigma 
+ \sum_{p\sigma} \left( \epsilon^{(l)}_{p\sigma} l_{p\sigma}^\dag l_{ p\sigma }
+\epsilon^{(r)}_{p\sigma} r_{p\sigma }^\dag r_{p\sigma } \right),
\end{equation}
where for the quantum dot we have assumed a single orbital level is relevant and have included the effect of the Zeeman term. The sign of $b=-g \mu_B B_z$ depends both on $g$ and $B_z$ but has no effect on our discussion so we fix $b > 0$. In Eq.~(\ref{H_ideal}), $l_{p\sigma}$ ($r_{p\sigma}$) destroys an electron in the state $p$ of the left (right) lead, with $\sigma=+(-)$ corresponding to $\uparrow(\downarrow)$. The single-particle energies $\epsilon^{(l,r)}_{p\sigma}$ are spin-dependent and determine the densities-of-states $\nu_{l\sigma},\nu_{r\sigma}$ at the Fermi levels $\mu_{l,r}$. Assuming identical leads, we have $\nu_{l \uparrow}=\nu_{r\downarrow}=\nu_+$ and $\nu_{l\downarrow}=\nu_{r\uparrow}=\nu_-$ for the majority and minority carriers, respectively. Furthermore, we consider spin-independent tunneling:
\begin{equation}
H_T = \sum_{p\sigma} t_l l_{p\sigma}^\dag d_\sigma +\sum_{p\sigma} t_r r_{p\sigma}^\dag d_{\sigma} + {\rm h.c.},
\end{equation}
where h.c. indicates hermitian conjugate terms. 

\begin{figure}[t]
\includegraphics[width=0.47\textwidth]{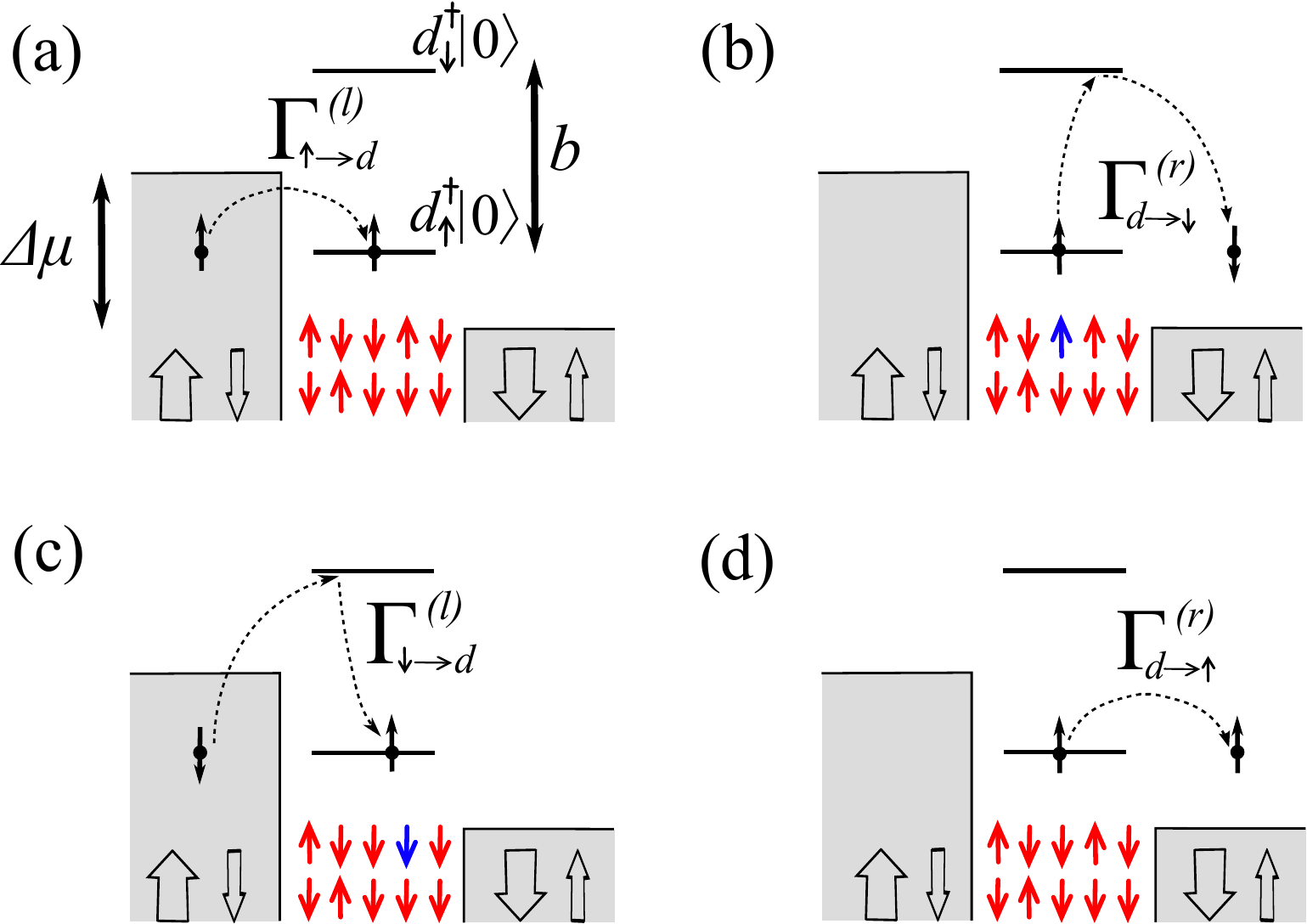}
\caption{\label{transport} Schematic of the four relevant tunneling processes through a quantum-dot spin valve with positive bias $\Delta \mu$ and large Zeeman splitting $b$. The polarizations of majority/minority carriers in the ferromagnetic leads are indicated with wide/narrow empty arrows. Panels (a) and (d) show direct tunneling events, with rates given by Eqs.~(\ref{tunnel_rate_noflip1}) and (\ref{tunnel_rate_noflip2}). Panels (b) and (c) show second-order tunneling events that involve an electron-nuclear spin flip-flop through hyperfine coupling. The corresponding rates are given in Eqs.~(\ref{tunnel_rate_flipL1}) and (\ref{tunnel_rate_flipR1}). Processes (c) and (d) are suppressed for half-metal leads. }
\end{figure}

The full Hamiltonian, including the nuclear bath, reads
\begin{equation}\label{Htot}
H= H_{\rm el}+ H_T + H_{\rm hf}+ H_{N},
\end{equation}
where $H_{\rm hf}$ is given by the two terms in Eq.~(\ref{hf}). The last term is an inhomogeneous field acting on each nuclear spin:
\begin{equation}\label{H_N}
H_N=\sum_i b_k I_{z,k},
\end{equation}
which includes the small Zeeman splitting of the nuclear spins ($b_k \ll b$) and accounts phenomenologically for all possible sources of inhomogeneity (e.g., spatial variations of the magnetic field $\bf B$, slightly non-uniform couplings $A_k$, or the nuclear-spin dipole-dipole interactions). 

In the following, we will be especially interested in electron-nuclear flip-flop processes. In fact, transport through the spin valve induces a non-trivial nuclear-spin dynamics through $H^{\rm hf}_{ff}$, defined in Eq.~(\ref{hf}). More specifically, it is convenient to assume zero temperature for the electron system (but not for the nuclear spins) and a relatively large magnetic field, such that only the $\uparrow$ state of the dot is in the bias window of width $\Delta\mu=\mu_l -\mu_r$ (see Fig.~\ref{transport}). For this arrangement, tunneling out of the dot is suppressed due to the reduced density-of-states for the minority-carrier band ($\uparrow$) in the right lead. In the extreme limit of half-metal leads ($\nu_-=0$), direct tunneling out of the dot becomes impossible and the current to the right lead is dominated by second-order processes involving both $H_T$ and $H^{\rm hf}_{ff}$, as illustrated in Fig.~\ref{transport}. Clearly, each electron passing through the spin valve will transfer $\hbar$ of angular momentum to the nuclear-spin system. Even with imperfectly-polarized ferromagnetic leads, such spin-flip processes induce a transfer of angular momentum to the nuclear spins. 

We also note that the left ferromagnet in Fig.~\ref{transport} is not strictly necessary for the transfer of spin, when a fixed bias direction is chosen. However, an advantage of the setup shown in Fig.~\ref{transport} is that nuclear spins can be polarized in either direction by reversing the bias ($\Delta\mu\to -\Delta\mu$). 

\section{Tunneling rates at large magnetic field} \label{sec_rates}

We now discuss the general form of the tunneling rates, which determine the electron transport through the spin valve. It is most transparent to work with sufficiently large magnetic field and in the weak tunneling limit, such that both $H_{\rm hf}$ and $H_T$ can be treated as perturbations. The precise condition on $b$ is given towards the end of this Section, together with some considerations on what to expect outside the large-$b$ parameter regime (based on the non-perturbative results of Sec.~\ref{exact_rates_half_metal}).

In the large-$b$ limit,  the unperturbed eigenstates are $|\Psi^{\rm el}_n\rangle \otimes |\Psi^{N}_m\rangle$, where $|\Psi^{\rm el}_n\rangle$ ($|\Psi_m^N\rangle$) are the electronic (nuclear) eigenstates of $H_{\rm el}$ ($H_N$), with $n$ ($m$) a suitable index. We also restrict ourselves to the case of forward bias, $\Delta \mu >0$, for which the four relevant transport processes are illustrated in Fig.~\ref{transport}. The corresponding tunneling rates are denoted $\Gamma^{(l)}_{\alpha \rightarrow d}$ and $\Gamma^{(r)}_{d \rightarrow \alpha}$, where we only specify the spin orientation ($\alpha=\uparrow/\downarrow$) in the right/left $(r/l)$ lead, since in the quantum dot ($d$) the spin is always $\uparrow$. The reverse-bias case, $\Delta \mu <0$, can be treated in a similar way, by considering the relevant rates $\Gamma^{(l)}_{\alpha \leftarrow d}$ and $\Gamma^{(r)}_{d \leftarrow \alpha}$.

The spin-conserving rates due to $H_T$ are (at vanishing temperature, $T=0$):
\begin{eqnarray}
&&\Gamma_{\uparrow \rightarrow d}^{(l)} = 2\pi \nu_+ |t_{l}|^2 \equiv \Gamma_+^{(l)}, \label{tunnel_rate_noflip1}\\
&&\Gamma_{d \rightarrow \uparrow}^{(r)} = 2\pi \nu_- |t_{r}|^2 \equiv \Gamma_-^{(r)}, \label{tunnel_rate_noflip2}
\end{eqnarray}
where the weak-tunneling condition requires $\Gamma_\pm^{(l,r)}\ll b$. As for the spin-flip tunneling rates, these involve necessarily both $H_T$ and $H^{\rm hf}_{ff}$, thus need to be computed with higher-order perturbation theory. By specializing to $\Gamma_{d \rightarrow \downarrow}^{(r)}$ and to an initial nuclear state $|\Psi^N_i\rangle$ (the initial electron state is $|\Psi^{\rm el}_i\rangle = d^\dag_\uparrow |0\rangle$) we have:
\begin{eqnarray}\label{2ndord_FGR}
\Gamma_{d \rightarrow \downarrow}^{(r)}=2\pi\sum_{f',f''} \left| \langle \Psi_{f'}^N | \langle \Psi_{f''}^{\rm el} | H_T  \frac{1}{H_{0}-E_i} H^{\rm hf}_{ff}|\Psi^{\rm el}_i\rangle |\Psi^N_i\rangle\right|^2 \nonumber   \\
\times \delta(E^{\rm el}_{f''}+E^N_{f'} -E_i), \hspace{2 cm}
\end{eqnarray}
where $E_i$ is the eigenvalue of $H_0=H_{\rm el}+H_N$ for the initial state and $E^{N}_{f'}$ ($E^{\rm el}_{f''} $) is the eigenvalue of $H_{N}$ ($H_{\rm el}$) for the final state. Since $H^{\rm hf}_{ff}$ flips the electron spin in the dot, we approximate $ H_{0}-E_i \simeq b $, which is applicable when $b_{N,i} \ll b$, thus the change in nuclear energy induced by $H^{\rm hf}_{ff}$ can be neglected. Then Eq.~(\ref{2ndord_FGR}) evaluates to:
\begin{equation}
\Gamma_{d \rightarrow \downarrow}^{(r)}=\eta \Gamma^{(r)}_+ \sum_{f'} \left| \langle \Psi_{f'}^N | I_+| \Psi^N_i\rangle\right|^2 = \eta \Gamma_+^{(r)} \langle \Psi_{i}^N |I_- I_+| \Psi^N_i\rangle,
\end{equation}
where we have introduced $\Gamma^{(r)}_+ =  2\pi \nu_+ |t_{r}|^2$ (similarly, $\Gamma^{(l)}_- =  2\pi \nu_- |t_{l}|^2$) and
\begin{equation}
\eta = \left(\frac{A}{2N b}\right)^2.
\end{equation}

The other spin-flip rate, $\Gamma_{\downarrow \rightarrow d }^{(l)}$, can be obtained with the same method. The results are immediately extended to a nuclear state which is an incoherent mixture of eigenstates, i.e., with density matrix $\rho_N = \sum_{i} p_i | \Psi^N_i \rangle \langle  \Psi^N_i |$, which will be useful in later sections. Finally, we can write the spin-flip rates as follows:
\begin{eqnarray}
&&\Gamma_{\downarrow \rightarrow d}^{(l)}[\rho_N] = \eta \, \Gamma^{(l)}_- \,{\rm Tr} [ I_+ I_- \rho_N ], \label{tunnel_rate_flipL1}\\
&&\Gamma_{d \rightarrow \downarrow}^{(r)}[\rho_N] = \eta \, \Gamma^{(r)}_+ \,{\rm Tr} [ I_- I_+ \rho_N ]. \label{tunnel_rate_flipR1}
\end{eqnarray}
The validity of Eq.~(\ref{2ndord_FGR}) requires the matrix element from the initial to the intermediate state to be smaller than $b$. By considering total angular momentum eigenstates $|I,m \rangle$ for the nuclear-spin system (where $m=-I,\ldots, I$ is the eigenvalue of $I_z$), the most restrictive condition is obtained for $|N/2, 0 \rangle$ (assuming spin-1/2 nuclei), giving $\frac{A}{2N}\langle \Psi_{f'}^N | I_+| \Psi^N_i\rangle \simeq A/4 \ll b$. The actual range of magnetic fields at which this condition is satisfied depends on the specific system, and can be quite accessible for group-IV materials. Using $A\simeq 4.3~\mu$eV and $g=2$, appropriate for a ${}^{29}$Si quantum dot,\cite{Assali2011} gives $B_z \gg 40$~mT. This requirement is further relaxed if the number of spin-carrying nuclei $N$ is only a fraction of $N_{\rm dot}$ (the total number of atoms within a quantum-dot Bohr radius). Then, $A$ should be rescaled by a factor $N/N_{\rm dot}<1$. As discussed in Sec.~\ref{sec:SiGenanowires}, this scenario is relevant to realize a uniform hyperfine coupling.

We also comment about the smallness of $\eta$ when $b \gg A$ and $N \gg 1$. To reassure the reader, we anticipate that in a suitable regime of long-range nuclear-spin coherence the factor ${\rm Tr} [ I_\pm I_\mp \rho_N ]$ is of order $N^2$. Furthermore, a derivation of transition rates beyond the perturbative result of Eqs.~(\ref{tunnel_rate_flipL1})--(\ref{tunnel_rate_flipR1}) is discussed in Sec.~\ref{exact_rates_half_metal} and applies when the local nuclear fields $b_k$ are constant. In this case, the eigenstates of the quantum dot are known exactly \cite{Khaetskii2003,Eto2004,Coish2007} and we briefly review this exact solution in Appendix~\ref{sec:exact_eigenstates}. We find that the non-perturbative result is qualitatively similar to the large-$b$ limit, but $A/b$ is replaced by a factor of order unity [see Eq.~(\ref{gamma_m_zero_b}) and related discussion].   

A crucial feature of Eqs.~(\ref{tunnel_rate_flipL1})--(\ref{tunnel_rate_flipR1}) is the nuclear-spin factor, ${\rm Tr} [ I_\mp I_\pm \rho_N ]$, since this results in a strong dependence of the tunneling rates on coherence properties of the nuclear-spin bath. This dependence becomes especially pronounced when contrasting the regime in which the total angular momentum $I$ is conserved with the opposite situation, in which a fast local nuclear-spin dephasing mechanism exists. The difference between these two limits is illustrated most clearly for the case of half-metal leads, which we discuss next.

\section{The half-metal limit} \label{sec_half_metal}

For half-metal leads ($\nu_- =0$) we assume that a fully-polarized nuclear-spin state can be prepared. Nuclear-spin dark states \cite{Imamoglu2003,Taylor2003} are an obstacle towards reaching full polarization but the pure dephasing term $H_N$ can induce transitions out of the dark states.\cite{Taylor2003} Thus, in our model, this term allows the system to reach the stationary state $|N/2, - N/2 \rangle$ (for spin-1/2 nuclei and $\Delta\mu<0$). After initialization, switching the bias to $\Delta \mu >0$ leads to
\begin{equation}\label{ideal_chain}
|\tfrac{N}{2},-\tfrac{N}{2} \rangle \to |\tfrac{N}{2},-\tfrac{N}{2}+1 \rangle \to |\tfrac{N}{2},-\tfrac{N}{2}+2 \rangle \to \ldots ,
\end{equation}
where we assume nuclear-spin dephasing is sufficiently weak that the total angular momentum is conserved during the time evolution. In particular, this assumption is justified if all the couplings $A_k,b_k$ in Eq.~(\ref{Htot}) are approximately constant.
In Sec.~\ref{half-metal-rates} below we also address the opposite case with strong local nuclear-spin dephasing. The transition rates can be immediately obtained at large $b$ from Eqs.~\eqref{tunnel_rate_flipL1}--\eqref{tunnel_rate_flipR1} but we will also discuss the rates for arbitrary $b$ in Sec.~\ref{exact_rates_half_metal}, making use of the exact solution given in Appendix~\ref{sec:exact_eigenstates}.

\subsection{Large magnetic field}\label{half-metal-rates}

For positive bias, the relevant spin-flip transition rates are Eqs.~(\ref{tunnel_rate_flipL1}) and (\ref{tunnel_rate_flipR1}). By making use of the angular-momentum states $|I, m\rangle$, these rates become:
\begin{eqnarray}
&&\Gamma_{\downarrow \rightarrow d}^{(l)}= \eta \, \Gamma^{(l)}_- (I+m)(I-m+1), \label{tunnel_rate_flip_coh1}\\
&&\Gamma_{d \rightarrow \downarrow}^{(r)} = \eta \, \Gamma^{(r)}_+ (I-m)(I+m+1). \label{tunnel_rate_flip_coh2}
\end{eqnarray}
This result shows that a large enhancement of the tunneling rate can be realized since we assume that full polarization can be reached for half-metal leads. Further assuming nuclear-spin $I_k=1/2$ for simplicity, we have $I=N/2$ for the collection of $N$ coupled nuclear spins. The largest enhancement is at $m=0$, when Eq.~(\ref{tunnel_rate_flip_coh2}) gives $\Gamma_{\downarrow \rightarrow d}^{(l)}\simeq \eta \, \Gamma^{(l)}_- N^2/4 $ (while $\Gamma_{d \rightarrow \downarrow}^{(r)}=0$, due to $\nu_-=0$). This enhancement is directly due to quantum coherence in the $|N/2,0\rangle$ nuclear state. On the other hand, the uncorrelated initial state at $m=-N/2$ gives $\Gamma_{\downarrow \rightarrow d}^{(l)}\simeq \eta \, \Gamma^{(l)}_- N $. The latter result is proportional to $N$, which is the expected dependence for a tunneling process where the electron spin flip-flop with the individual nuclear spins occurs incoherently.

To illustrate this coherent enhancement more clearly, we consider an incoherent mixture of `product states' of type $|n\rangle = |\uparrow \downarrow \downarrow \uparrow\downarrow \ldots \rangle$ and fixed value of $m$. In this case, Eqs.~(\ref{tunnel_rate_flipL1}) and (\ref{tunnel_rate_flipR1}) give (for generic imperfectly-polarized ferromagnetic leads with $\nu_- \neq 0$),
\begin{eqnarray}
&&\Gamma_{\downarrow \rightarrow d}^{(l)} = \eta \, \Gamma^{(l)}_- (N/2 + m), \label{tunnel_rate_flip_incoh1}\\
&&\Gamma_{d \rightarrow \downarrow}^{(r)} = \eta \, \Gamma^{(r)}_+ (N/2 - m). \label{tunnel_rate_flip_incoh2}
\end{eqnarray}
These rates are appropriate if dephasing mechanisms of the nuclear-spin system act quickly on the time scale of individual tunneling events. This dephasing will occur, e.g., in the presence of strong variations in the $b_k$ of Eq.~(\ref{H_N}). In particular, the relevant rate $\Gamma_{d \rightarrow \downarrow}^{(r)}$ for half-metal leads is simply proportional to the number of $\downarrow$ nuclei, which allow the spin-flip tunneling process out of the dot. 

The enhancement factor of order $N$ is especially significant since a typical quantum dot has $N \sim 10^5 - 10^6$. However, the effect of nuclear-spin coherence can already be seen in the first few individual tunneling events at $I_z \simeq -N/2$. Since $\Gamma_{\uparrow \rightarrow d}^{(l)} \gg \Gamma_{d \rightarrow \downarrow}^{(r)} $, the rates $\gamma_m$ for the $I_z= m \to m+1$ transition of Eq.~(\ref{ideal_chain}) are approximated very well by Eq.~(\ref{tunnel_rate_flip_coh2}):
\begin{equation}\label{rates_chain_superr}
\gamma_{m} \simeq \eta N \Gamma_+^{(r)} , ~ 2 \eta N \Gamma_+^{(r)}, ~ 3 \eta N \Gamma_+^{(r)}, \ldots
\end{equation}
for $m+N/2= 0,1,2,\ldots$. These tunneling rates should be compared with the approximately constant rate $\eta N\Gamma_+^{(r)}$ for an incoherent nuclear-spin bath [see Eq.~(\ref{tunnel_rate_flip_incoh2}) with $m \sim -N/2$]. Thus, if single tunneling events can be  detected,\cite{Pekola2013} observing $\gamma_{m+1}/\gamma_{m} >1 $ provides a signature of nuclear-spin coherence. The difference is largest for the first few tunneling events, e.g., $(\gamma_{-\frac{N}{2}+1})/( \gamma_{-\frac{N}{2}})\simeq 1$ for an incoherent nuclear-spin mixture and $(\gamma_{-\frac{N}{2}+1})/( \gamma_{-\frac{N}{2}}) \simeq 2$ in Eq.~(\ref{rates_chain_superr}) above (while $\gamma_{m+1}/\gamma_{m}\to 1$ for larger $m$ also in the coherent case). 
 
The enhancement of spin-flip rates through nuclear-spin coherence in such a spin valve setup is analogous to the enhancement analyzed in Ref.~\onlinecite{Eto2004} in the spin-blockade regime of a double quantum dot. There, it was noted that for uniform hyperfine couplings the transport becomes analogous to the superradiant emission of an ensemble of $N$ two-level atoms.\cite{Dicke1954,Gross1982} An alternative transport setup based on this analogy has also been recently examined in Ref.~\onlinecite{Schuetz2012}. Thus, by interpreting the $\uparrow/\downarrow$ nuclear-spin states as ground/excited atomic states, a spin-flip tunneling event corresponds to the emission (or absorption) of a photon. Given $N$ initially excited atoms, each with an independent decay rate $\eta\Gamma_+^{(r)}$, Eq.~(\ref{rates_chain_superr}) is then immediately understood as the decay rate due to collective photon emission.\cite{Gross1982} The `superradiant' regime can be contrasted with the photon emission rate $\eta \Gamma_+^{(r)}N_\downarrow$ for the usual spontaneous emission of $N_\downarrow$ excited atoms (with $N_\downarrow = N/2 - m$), corresponding to an incoherent nuclear-spin bath. 

\subsection{Rates from exact electron-nuclear spin eigenstates}\label{exact_rates_half_metal}

We can extend the previous results to arbitrary magnetic field using the exact eigenstates described in Appendix~\ref{sec:exact_eigenstates}. In direct analogy with Eq.~(\ref{ideal_chain}), and assuming that only the lower energy level $\epsilon_{I,m}^-$ is in the bias window, we can write the evolution of the coupled electron-nuclear-spin system as  
\begin{equation}\label{tunneling_exact}
\ldots \to |I,m \rangle |0\rangle \to \left|\varphi^-_{I}(m)\right> \to |I,m+1 \rangle |0\rangle \to \ldots .
\end{equation}
The corresponding rates are obtained from Eq.~\eqref{exact_eigenstates2} as:
\begin{align}
\Gamma_{\uparrow \rightarrow d}^{(l)} &= \Gamma_+^{(l)} |\beta_{I,m}|^2, \label{exact_rates_in_plus}\\
\Gamma_{d \rightarrow \downarrow}^{(r)}&= \Gamma_+^{(r)} |\alpha_{I,m}|^2. \label{exact_rates_out_plus}
\end{align}
From Eq.~(\ref{coefficients_large_b}), we immediately recover the results from the perturbative treatment at large $b$, since Eq.~(\ref{exact_rates_in_plus}) yields Eq.~(\ref{tunnel_rate_noflip1}) while Eq.~(\ref{exact_rates_out_plus}) yields Eq.~(\ref{tunnel_rate_flip_coh2}). On the other hand, these expressions allow us to go beyond the perturbative limit and discuss the opposite case, $b\ll A$. Assuming a single energy level, $\epsilon_{N/2,m}^-$, in the bias window, and using Eq.~(\ref{coefficients_small_b}), we obtain:
\begin{align}
\Gamma_{\uparrow \rightarrow d}^{(l)} &= \Gamma_+^{(l)} \frac{N/2-m}{N+1}, \label{exact_rates_in}\\
\Gamma_{d \rightarrow \downarrow}^{(r)}&= \Gamma_+^{(r)} \frac{N/2+m+1}{N+1}, \label{exact_rates_out}
\end{align}
where, once again, we have assumed the fully-polarized initial condition, $I=N/2$. In this case, the hypothesis of a single energy level $\epsilon_{N/2,m}^-$ in the bias window can be easily satisfied due to the relatively large energy splitting $\epsilon_{N/2,m}^+-\epsilon_{N/2,m}^-\simeq A/2$, which, as seen in Eq.~(\ref{energies_large_b}), is independent of $m$ when $b=0$.

To discuss the enhancement of electronic current through the spin valve due to nuclear-spin coherence, we compute the rate for the process $|N/2,m \rangle \to |N/2,m+1 \rangle $, given by $\gamma_m=( 1/\Gamma_{\uparrow \rightarrow d}^{(l)} + 1/\Gamma_{d \rightarrow \downarrow}^{(r)})^{-1}$. By assuming for simplicity $\Gamma_+^{(l)}=\Gamma_+^{(r)}$, we obtain:
\begin{equation}\label{gamma_m_zero_b}
\gamma_m = \Gamma_+^{(r)}\frac{(N/2-m)(N/2+m+1)}{(N+1)^2},
\end{equation}
which has the same form as Eq.~(\ref{tunnel_rate_flip_coh2}) with the substitution $\eta \to (N+1)^{-2}$. Thus, all previous remarks on the enhancement of the coherent rates at $I_z=0$ still hold in this case. While decreasing $b$ clearly enhances the prefactor $\eta =A^2/(2Nb)^2$ in Eq.~(\ref{tunnel_rate_flip_coh2}), the perturbative treatment fails at small $b$ and the tunneling rate saturates to the value given in Eq.~(\ref{gamma_m_zero_b}). For $m=0$, Eq.~(\ref{gamma_m_zero_b}) gives $\gamma_m \simeq \Gamma_+^{(r)}/4$, comparable to the direct tunneling rate in the absence of any electron-nuclear spin-flip mechanisms.

\section{Dynamics with imperfectly-polarized ferromagnetic leads}\label{sec_nuclear_evolution}

We now return to the limit of large $b$ and discuss the general case of ferromagnetic leads with imperfect polarization. In general, the evolution of the reduced nuclear-spin density matrix $\rho_N$ is described by a quantum master equation. This approach will be described in Sec.~\ref{sec_mastereq}, where we will also justify in more detail how, in the two limiting regimes of very fast/slow nuclear-spin decoherence, the nuclear-spin dynamics can be characterized through much simpler rate equations. As discussed here, the final result can be derived more physically directly from the transition rates discussed in the previous section.

We start from the incoherent evolution, for which we suppose a nuclear-spin state of the form
\begin{equation}\label{rho_m_def}
\rho_N(t) =\sum_{m} p_{m}(t) \sum_{n_m} \frac{|n_m \rangle \langle n_m|}{\binom{N}{N/2+m}} = \sum_{m} p_{m}(t) \rho_{m},
\end{equation}
where $|n_m\rangle$ is a complete basis of eigenstates of $I_z$. As seen above, here $\rho_{m}$ is maximally mixed in the $I_z=m$ subspace, which is justified in the presence of fast local nuclear-spin dephasing. Transitions $m \to m \pm 1$ in the nuclear-spin system are induced by electron tunneling events with the rates obtained in Sec.~\ref{sec_rates}, such that the $p_m$ satisfy:
\begin{equation}\label{rho_ferromagnet_rate_eq}
\dot{p}_m = \gamma_{m-1}^+ p_{m-1}+ \gamma_{m+1}^- p_{m+1} - (\gamma_{m}^+ +\gamma_{m}^-)p_m.
\end{equation}
To find the value of $\gamma_m^\pm$, we use the fact that the spin-flip tunneling rates are small: $\Gamma_{\downarrow \to d}^{(l)} ,\Gamma_{d \to \downarrow}^{(r)} \ll \Gamma_{\uparrow \to d}^{(l)}, \Gamma_{d \to \uparrow}^{(r)}$. It then follows that, at any given moment, the quantum-dot occupation $n_d$ is approximately determined by the direct tunneling processes:
\begin{equation}\label{n0_I0_ferrom}
n_d \simeq \frac{\Gamma_{+}^{(l)}}{\Gamma_{+}^{(l)}+\Gamma_{-}^{(r)}}.
\end{equation}
Given the probability $n_d$ that the dot is full, a nuclear-spin-flip process $I_z = m \to m+1$ can occur due to a flip-flop tunneling event from the quantum dot to the right lead:
\begin{equation}\label{gamma_plus_def}
\gamma_m^+ = n_d ~ \Gamma_{d \to \downarrow}^{(r)}[\rho_{m}] = \eta \Gamma_+ (N/2 - m),
\end{equation}
while the rate for $m \to m-1$ is determined by a flip-flop tunneling event from the left lead to the empty quantum dot:
\begin{equation}\label{gamma_minus_def}
\gamma_m^- = (1-n_d)~\Gamma_{\downarrow \to d}^{(l)}[\rho_{m}] = \eta \Gamma_- (N/2 + m).
\end{equation}
Equations (\ref{gamma_plus_def}) and (\ref{gamma_minus_def}) are computed here using Eqs.~(\ref{tunnel_rate_flipL1}) and (\ref{tunnel_rate_flipR1}) with the fully mixed states $\rho_{m}$ given in Eq.~(\ref{rho_m_def}). We have also introduced:
\begin{equation}\label{Gamma_pm_def}
\Gamma_\pm = \frac{\Gamma_{\pm}^{(l)}\Gamma_{\pm}^{(r)}}{\Gamma_{+}^{(l)}+\Gamma_{-}^{(r)}}.
\end{equation} 

We now consider the limit of negligible nuclear-spin dephasing, for which a suitable choice of $\rho_N$ is in terms of total angular-momentum eigenstates
\begin{equation}\label{rho_def_coh}
\rho_N(t) =\sum_{I,m} p_{I,m}(t) |I, m \rangle \langle I, m|.
\end{equation}
For simplicity, in Eq.~(\ref{rho_def_coh}) we have omitted a sum over the permutation index,\cite{Arecchi1972} which does not enter the transition rates. The degeneracies will be accounted for appropriately when needed. Because of the uniform hyperfine interaction, the flip-flop tunneling processes conserve $I$ but induce nuclear spin-flip transitions $m \to m \pm 1$. The relevant tunneling rates are obtained as in Eqs.~(\ref{gamma_plus_def}) and (\ref{gamma_minus_def}) but with states $|I, m \rangle \langle I, m|$ instead of $\rho_m$, 
\begin{equation}\label{gamma_coh_def}
\gamma_{I,m}^\pm = \eta \Gamma_\pm (I \mp m)(I \pm m +1),\quad{\rm (coherent)}
\end{equation}
The time evolution of $p_{I,m}$ follows from Eq.~\eqref{gamma_coh_def} as
\begin{align}\label{rho_ferromagnet_rate_eq_coh}
\dot{p}_{I,m} = &\gamma_{I,m-1}^+ p_{I,m-1}+ \gamma_{I,m+1}^- p_{I,m+1} \nonumber \\
& - (\gamma_{I,m}^+ +\gamma_{I,m}^-)p_{I,m}.
\end{align}
As seen in Eq.~\eqref{rho_ferromagnet_rate_eq_coh} above, it is simple to recover the limit of half-metal leads discussed in the previous section. In that case, we have $\Gamma_+ = \Gamma_{+}^{(r)}$, $\Gamma_- = 0$ and maximal angular momentum $I=N/2$. In the half-metal limit, Eq.~(\ref{gamma_coh_def}) recovers the result of Eq.~(\ref{rates_chain_superr}), $\gamma_{N/2,m}^+ =\gamma_m$ and $\gamma_{N/2,m}^- =0$. We also note that the probability
$P(I)=\sum_{m} p_{I,m}$ is independent of time due to angular-momentum conservation, which is straightforward to verify from Eq.~(\ref{rho_ferromagnet_rate_eq_coh}).  In addition, we define $p_m(t)$ as follows
\begin{equation}\label{P_I_def}
p_m(t) = \sum_I p_{I,m}(t) = \sum_{I} P(I) p_{m|I}(t).
\end{equation} 
Equation \eqref{P_I_def} can be compared more readily to the incoherent result of Eq.~(\ref{rho_ferromagnet_rate_eq}). In Eq.~\eqref{P_I_def} we have introduced the conditional probability $p_{m|I}(t)=p_{I,m}/P(I)$. Since $P(I)$ is time-independent in the coherent regime, $p_{m|I}(t)$ obeys the same equation of motion as $p_{I,m}(t)=p_{m|I}(t)P(I)$, see Eq.~(\ref{rho_ferromagnet_rate_eq_coh}).

\subsection{Stationary states} \label{sec_stationary}
Before discussing dynamics of the nuclear-spin magnetization, here we first analyze the stationary solution of Eq.~(\ref{rho_ferromagnet_rate_eq}), which can be found directly from the analogy with a system of $N$ 2-level atoms discussed at the end of Sec.~\ref{sec_rates}. For independent atoms, the probabilities of occupying the $\uparrow/\downarrow$ states are $\Gamma_\pm/(\Gamma_+ + \Gamma_-)$.  These populations can be interpreted in terms of a fictitious spin temperatures $T_k^*$ at each nuclear site:
\begin{equation}\label{T_fictitious}
k_B T_k^*=\Delta_k \left(\ln \frac{\Gamma_-}{\Gamma_+} \right)^{-1}=\frac{\Delta_k}{\ln R}.
\end{equation}
where $\Delta_k$ is the energy splitting for nucleus $k$. A possible choice for $\Delta_k$ is its time-averaged value (i.e., $\Delta_k =|b_k +n_d A/2N|$ if $A/N$ is sufficiently small). However, any redefinition of $\Delta_k$ can be absorbed into the definition of $T_k^*$, the relevant parameter being $\Delta_k/T_k^*$ or, equivalently,
\begin{equation}
R=\frac{\Gamma_-}{\Gamma_+}= \left(\frac{\nu_-}{\nu_+}\right)^2.
\end{equation}
The stationary nuclear polarization is then given in terms
of a binomial distribution,  
\begin{equation}\label{stationary_incoherent_ferrom}
p_m^{\rm eq} =\binom{N}{N/2+m}\frac{R^{N/2-m}}{(1+R)^N}, \quad {\rm(incoherent)}
\end{equation}
which satisfies Eq.~(\ref{rho_ferromagnet_rate_eq}), as can be easily checked. For large $N$, Eq.~(\ref{stationary_incoherent_ferrom}) is a narrow distribution with
\begin{equation}\label{m_avg_incoh}
\langle m \rangle = \frac{N}{2}\left(\frac{1-R}{1+R}\right), \quad\langle \Delta m^2 \rangle = \frac{N R}{(1+R)^2},
\end{equation}
where $\langle m \rangle =\sum_m m p_m$ indicates the average with respect to the nuclear-spin distribution and $\Delta m = m-\langle m \rangle$. 

The coherent rates given in Eq.~(\ref{gamma_coh_def}) can also be interpreted in terms of a single fictitious temperature $T^*$ since we have $\gamma^-_{I,m} /\gamma^+_{I,m-1}=\Gamma_-/\Gamma_+ = e^{-\Delta/k_B T^*} $, independent of $I$ and $m$. $T_k^*=T^*$ independent of $k$ in the homogeneous limit [$\Delta_k=\Delta$ in Eq.~(\ref{T_fictitious})], required for the coherent regime. However, in the coherent regime, the physical picture is quite different since Eq.~(\ref{rho_ferromagnet_rate_eq_coh}) describes the thermalization of a $(2I+1)$-level system with constant energy-level spacing $\Delta$. For a given $I$, the equilibrium probability for occupation of state $|I, m\rangle$ is $\propto e^{-(I-m) \Delta/k_B T^*}$, which gives 
\begin{equation}\label{coherent_initial_ferromagn}
p_{m|I}^{\rm eq} = \frac{(1-R)R^{I-m}}{1-R^{2I+1}}. \quad {\rm(coherent)}
\end{equation}
Equation~(\ref{coherent_initial_ferromagn}) is very narrow and almost fully polarized within a subspace of fixed $I$, with $\langle m \rangle \simeq I $ and $\langle \Delta m^2\rangle \simeq R/(1 - R)^2$. This distribution is thus very different from Eq.~(\ref{stationary_incoherent_ferrom}). However, in general we should also account for the distribution of angular momentum $P(I)$, see Eq.~(\ref{P_I_def}). 

\begin{figure}
\includegraphics[width=0.45\textwidth]{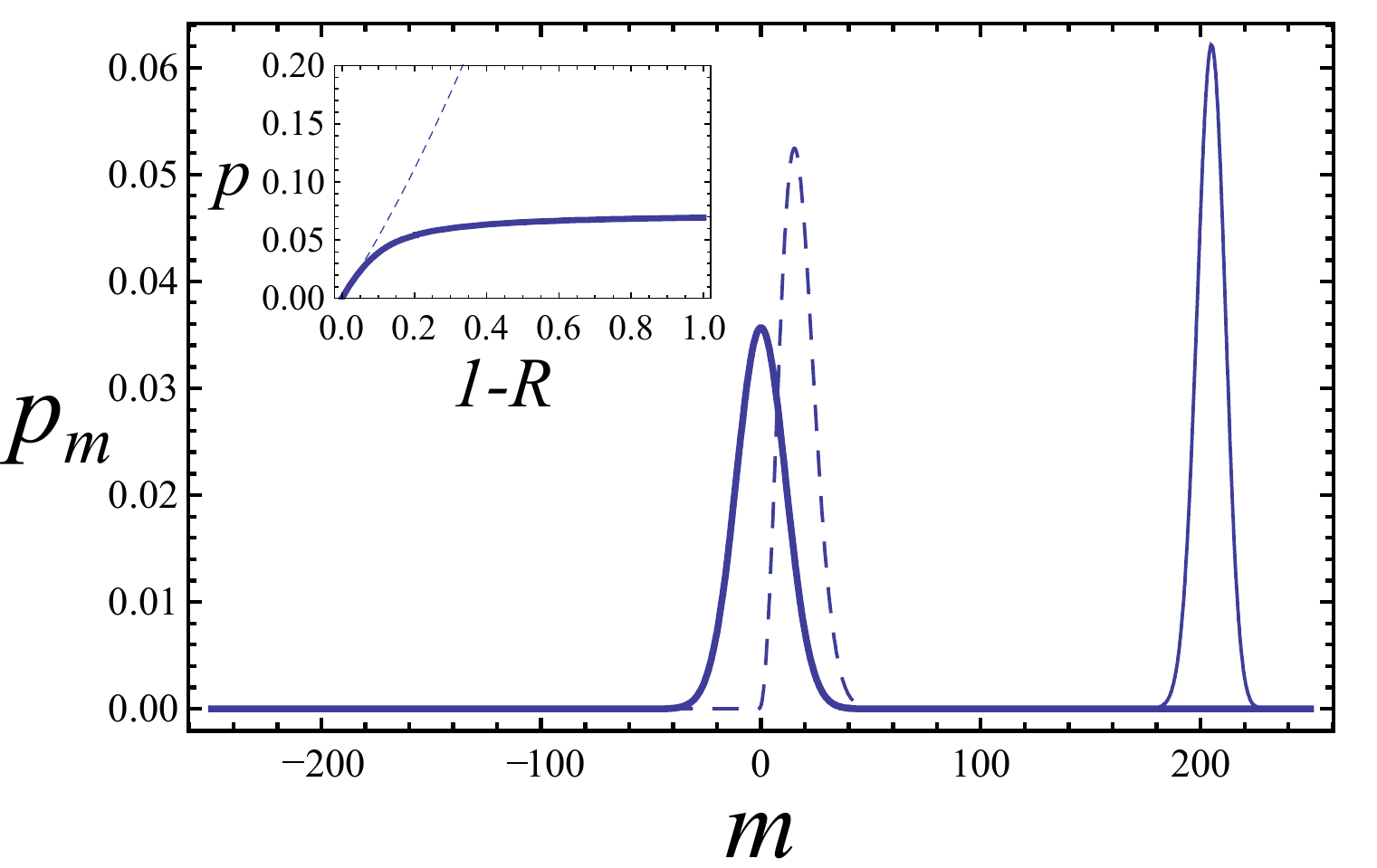}
\caption{\label{ferro_stationary} Stationary distributions of $p_m$, with $N=500$ and $R=0.1$. The thicker curve (with $\langle m \rangle =0$) is $p_m$ for a fully-mixed nuclear-spin state, assumed to be the initial state of the coherent evolution. The thinner dashed curve is the stationary state in the coherent regime. The thinner solid curve is the incoherent stationary state given by Eq.~(\ref{stationary_incoherent_ferrom}) and is actually independent of the initial state. The inset shows the dependence of the stationary polarization $p =2\langle m \rangle/N$ on $1-R$ in the coherent (solid) and incoherent (dashed) regimes. }
\end{figure}

As a first example, we consider a quantum dot disconnected from the ferromagnetic reservoirs, in which the nuclear-spin system has relaxed to a fully mixed state. Since the number of states with the same value of $I$ and $m$ is $D(N,I)=\binom{N}{N/2+I}-\binom{N}{N/2+I+1}$, we have:
\begin{equation}\label{P_I_fully_mixed}
P(I)= 2^{-N} D(N,I)(2I+1). 
\end{equation}
By allowing transport through the spin valve ($t_{l,r} \neq 0$) and assuming coherent nuclear-spin evolution, Eqs.~(\ref{P_I_def}) and (\ref{coherent_initial_ferromagn}) give the resulting stationary state. The limiting case $R \to 1$ (normal leads) results in a maximally mixed state, $p_m \to 2^{-N} \binom{N}{N/2+m}$.  This state is the same as the result for the incoherent distribution, Eq.~(\ref{stationary_incoherent_ferrom}), and is represented by the thick curve in Fig.~\ref{ferro_stationary}. Thus, unpolarized leads generally do not modify this initial distribution. For $R \to 0$ the maximum polarization, $m=I$, is reached in each subspace of fixed $I$.  In this case, Eq.~(\ref{P_I_def}) evaluates to $p_{m\geq 0} =P(m)$ and $p_{m<0}=0$. As is made clear from Fig.~\ref{ferro_stationary}, the two stationary distributions (coherent/incoherent) with the same $R$ are generally very different and, in particular, the nuclear polarization is much smaller for the coherent evolution. This difference in polarization is because, in the fully-disordered nuclear-spin state, the large majority of states have $I \simeq 0$. Thus, in the absence of an efficient relaxation mechanism for $I$, a large nuclear polarization cannot be achieved.\cite{Imamoglu2003,Taylor2003}

An expression can be found for $P(I)$ from the stationary state under incoherent evolution given in Eq.~(\ref{stationary_incoherent_ferrom}). Since Eq.~(\ref{rho_m_def}) describes a fully-mixed state within each $I_z=m$ subspace, the conditional probability is simply given by $P(I | m)= D(N,I) / \binom{N}{N/2+m}$, if $I \geq |m|$. Thus, in this case we obtain
\begin{align}\label{P_I_incoh}
P(I) & = \sum_{m=-I}^I P(I |m) p^{\rm eq}_m \nonumber\\
& = \frac{D(N,I) (1-R^{2I+1}) R^{N/2-I}}{(1-R)(1+R)^N}.
\end{align}
This distribution can be realized if a finite bias $\pm\Delta\mu$ is applied across the spin valve for a sufficiently long time, exceeding the nuclear-spin coherence time. We note that $p^{\rm eq}_m \to p^{\rm eq}_{-m} $ does not affect the distribution of $I$. Thus, it is possible to initialize the system at $\Delta \mu < 0$ with the incoherent distribution, Eq.~(\ref{P_I_incoh}), and to examine coherent dynamics upon reversing the bias. Evaluating Eq.~(\ref{P_I_def}) with Eqs.~(\ref{P_I_incoh}) and (\ref{coherent_initial_ferromagn}) gives the same $p^{\rm eq}_m$ previously given in Eq.~(\ref{stationary_incoherent_ferrom}). We thus arrive at the interesting result that the coherent evolution, combined with this $P(I)$ in Eq.~(\ref{P_I_incoh}), leads to the same stationary distribution as in the incoherent case.

This conclusion should not be very surprising. Physically, if full decoherence has taken place and the nuclear-spin system has reached a stationary state, this state will also be stationary for shorter time scales (at which it is justified to neglect decoherence). The analogy with a fictitious thermalization process at temperature $T^*$ offers an alternative point of view: while the coherent evolution leads to a fast thermalization process within each subspace with fixed $I$, it does not allow for thermalization between sectors with different $I$. On the other hand, the incoherent evolution leads to full thermalization. At long times, thermal equilibrium is established globally and within each $I$ sector, such that the final state is stationary for the coherent evolution as well. 

\subsection{Evolution of the nuclear magnetization} \label{sec_rate_eq_solutions}

We turn now to the dynamics of the nuclear magnetization. For the incoherent case, an exact solution of Eq.~(\ref{rho_ferromagnet_rate_eq}) which is a generalization of Eq.~(\ref{stationary_incoherent_ferrom}) can be found:
\begin{equation}
p_m(t)=\binom{N}{N_\uparrow} \frac{[\tilde\Gamma_+(t)]^{N_\uparrow}[\tilde\Gamma_-(t)]^{N_\downarrow}}{(\Gamma_+ + \Gamma_-)^N}, 
\end{equation}
with $N_{\uparrow/\downarrow} = N/2 \pm m$ and
\begin{equation}
\tilde\Gamma_\pm (t)=\Gamma_\pm \mp \Gamma_0 e^{-\eta(\Gamma_++\Gamma_-)t},
\end{equation} 
where $\Gamma_0$ determines the initial condition. Apart from the obvious remark that $\Gamma_0 = 0$ or sufficiently large $t$ yield back the stationary solution of Eq.~(\ref{stationary_incoherent_ferrom}), it is also interesting to consider $\Gamma_0 = \Gamma_+ - \Gamma_-$. This initial condition is the stationary state for negative bias $\Delta \mu$, and thus represents a natural starting point for the evolution of the nuclear-spin ensemble. The resulting nuclear magnetization reads
\begin{equation}\label{m_time_incoh}
\langle m \rangle = \frac{N}{2}\frac{1-R}{1+R}\left[ 1-2 e^{-\eta \Gamma_+ (1+R)t}\right].
\end{equation}

In the half-metal limit $\Gamma_+=\Gamma$ and $\Gamma_-=0$, the coherent dynamics is easily understood in terms of the optical analogy to superradiance. In fact, Eq.~(\ref{rho_ferromagnet_rate_eq_coh}) gives
\begin{equation}\label{rate_eq_superradiant}
\dot p_{I,m} =\gamma_{I,m-1} p_{I,m-1} - \gamma_{I,m} p_{I,m}, 
\end{equation}
with
\begin{equation}\label{rate_superradiant}
\gamma_{I,m}= \eta \Gamma (I-m)(I+m+1).
\end{equation}
This decay rate in the superradiant regime is very well known in the quantum optics community (see Ref.~\onlinecite{Gross1982} for a review). At a given value of $I$, the analytical solution for the distribution governing the nuclear magnetization is\cite{Gross1982}
\begin{equation}\label{rho_supperadiant}
p_{m|I}(t) \simeq \frac{4I^2 e^{-2\eta\Gamma I t}}{(I-m)^2} \exp\left[-\frac{2I (I+m)}{I-m} e^{-2\eta\Gamma I t} \right],
\end{equation}
for the initial condition $|I,I_z=-I \rangle$, large $I \gg 1$, and $t >1/(\Gamma I)$. Since half-metal leads would allow full polarization of the nuclear-spin bath, in this case it is justified to simply set $p_{I,m}= p_{m|(N/2)} \delta_{I, N/2} $, giving a complete description of the nuclear-magnetization dynamics. 

\begin{figure}
\includegraphics[width=0.45\textwidth]{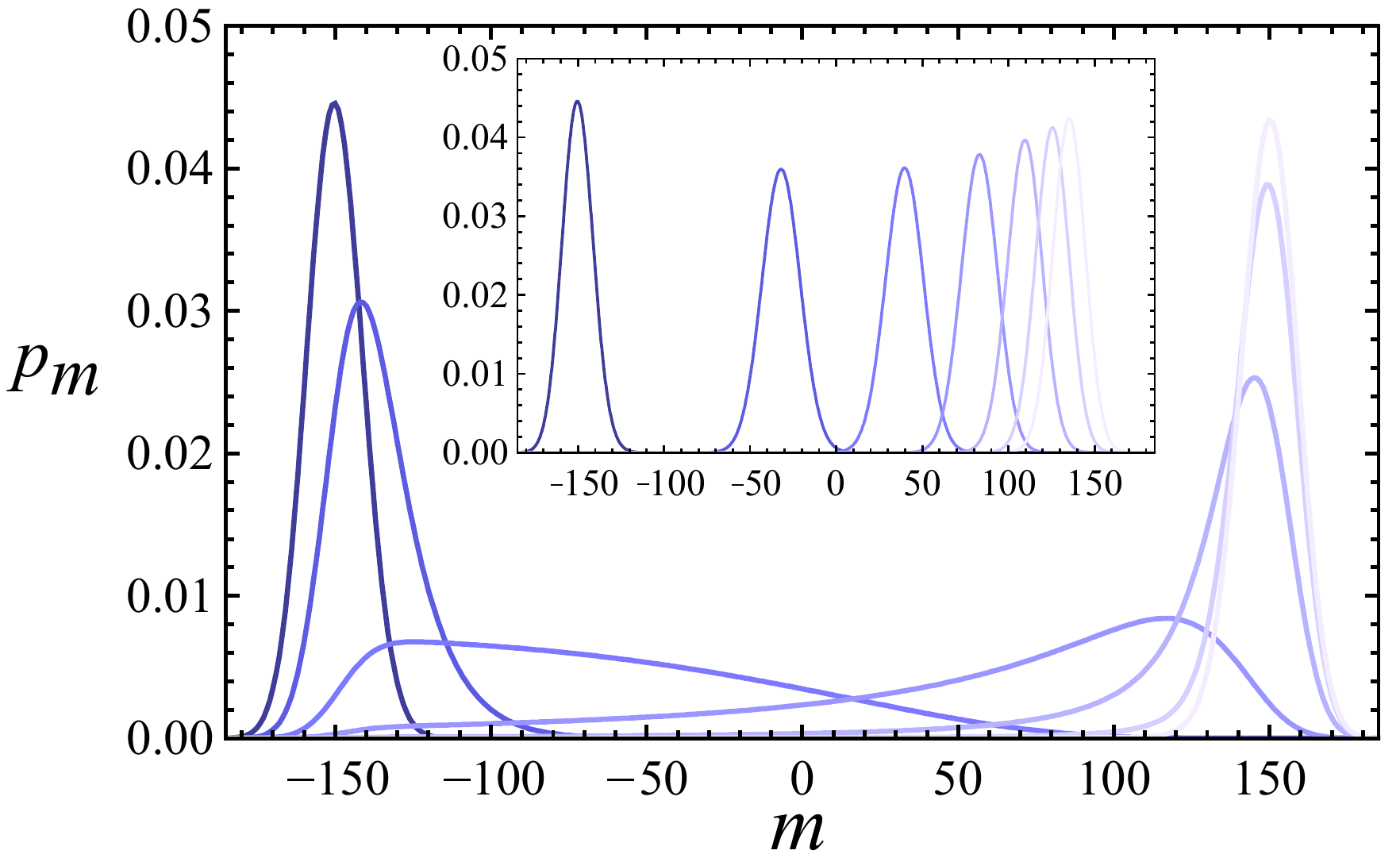}
\caption{\label{ferro_coherent_timeevol} Time evolution of $p_m$ in the coherent and incoherent regimes. Main panel: coherent time evolution of $p_m$ with $N=500$ and $R=0.25$. We plot the distributions at times $t_n=n \Delta t$, with $\eta \Gamma_+ \Delta t =0.01$ and $n=0,1,\ldots, 6$ (from darker to lighter color). At $t=0$ the distribution is $p^{\rm eq}_{-m}$ [see Eq.~(\ref{stationary_incoherent_ferrom})] and is close to $p^{\rm eq}_{m}$ at $n=6$. The inset shows the same result for incoherent dynamics and a larger time step $\eta \Gamma_+ \Delta t =0.4$.}
\end{figure}

The general case of imperfectly-polarized ferromagnetic leads cannot be mapped exactly to this superradiant description since Eq.~(\ref{rho_ferromagnet_rate_eq_coh}) is not of the form of Eq.~(\ref{rate_eq_superradiant}). Furthermore, the initial state is generally not $|I, -I \rangle$.  Instead, it is necessary to consider a mixture of different values of $I$ and $m$. We take the initial condition to be the stationary state at negative bias, $\Delta \mu<0$. In this case, the initial values of $p_{I,m}$ are simply obtained from Eq.~(\ref{stationary_incoherent_ferrom}) as $P(I|m)p^{\rm eq}_{-m}$, where $P(I|m)=D(N,I)/\binom{N}{N/2+m}$. We then numerically solve the simultaneous equations, Eq.~(\ref{rho_ferromagnet_rate_eq_coh}). An example of the resulting time evolution is shown in Fig.~\ref{ferro_coherent_timeevol} where, in contrast with the incoherent dynamics (inset), typical features of the superradiant behavior are recognized.  These features include a much faster dynamics (the timescale is shorter by a factor $\sim 40$) and the broad distribution at intermediate times ($\Delta m$ becomes of order $N$).  To make this connection explicit through an analytical treatment, we start from the approximation:
\begin{equation}\label{approx_rates}
\gamma_{I,m+1}^-  p_{I,m+1} - \gamma_{I,m}^-  p_{I,m}  \simeq \gamma_{I,m}^-  p_{I,m} - \gamma_{I,m-1}^-  p_{I,m-1}, 
\end{equation}
which is valid when $\gamma_{I,m+1}^-  p_{I,m+1} - \gamma_{I,m}^-  p_{I,m} $ has a weak dependence $m$ (e.g., if $\gamma_{I,m}$, $p_{I,m}$ are sufficiently broad functions of $m$). By using Eq.~(\ref{approx_rates}), we can rewrite the time evolution Eq.~(\ref{rho_ferromagnet_rate_eq_coh}) in the following form:
\begin{align}\label{approx_rate_eq_ferro}
\dot p_{I,m} \simeq (\gamma_{I,m-1}^+ -\gamma_{I,m-1}^- ) p_{I,m-1} - (\gamma_{I,m}^+ -\gamma_{I,m}^- ) p_{I,m}.
\end{align}
For $I \pm m \gg 1$, we can approximate the above rates as $\gamma_{I,m}^{\pm} \simeq \eta \Gamma_\pm (I^2 - m^2)$ or, with the same accuracy, 
\begin{equation}\label{approx_rates_ferro}
\gamma_{I,m}^+ -\gamma^-_{I,m} \simeq \eta(\Gamma_+ - \Gamma_-) (I-m)(I+m+1),
\end{equation}
which leads to a dynamics of the same form of Eqs.~(\ref{rate_eq_superradiant}) and (\ref{rate_superradiant}) and allows us to identify
\begin{equation}\label{Gamma_approx_superradiant}
\Gamma \simeq \Gamma_+ - \Gamma_-. 
\end{equation}
To obtain the appropriate value of $I$, we observe that for an incoherent initial state $p^{\rm eq}_{-m}$, Eq.~(\ref{P_I_incoh}) gives a narrow distribution $P(I)$ centered around
\begin{equation}\label{I_approx_superradiant}
\langle I \rangle \simeq -\langle m\rangle \simeq \frac{N}{2} \frac{1-R}{1+R}.
\end{equation}
We can thus assume that the initial state approximates $|I, -I \rangle $, with $I$ as in Eq.~(\ref{I_approx_superradiant}). 

\begin{figure}
\includegraphics[width=0.45\textwidth]{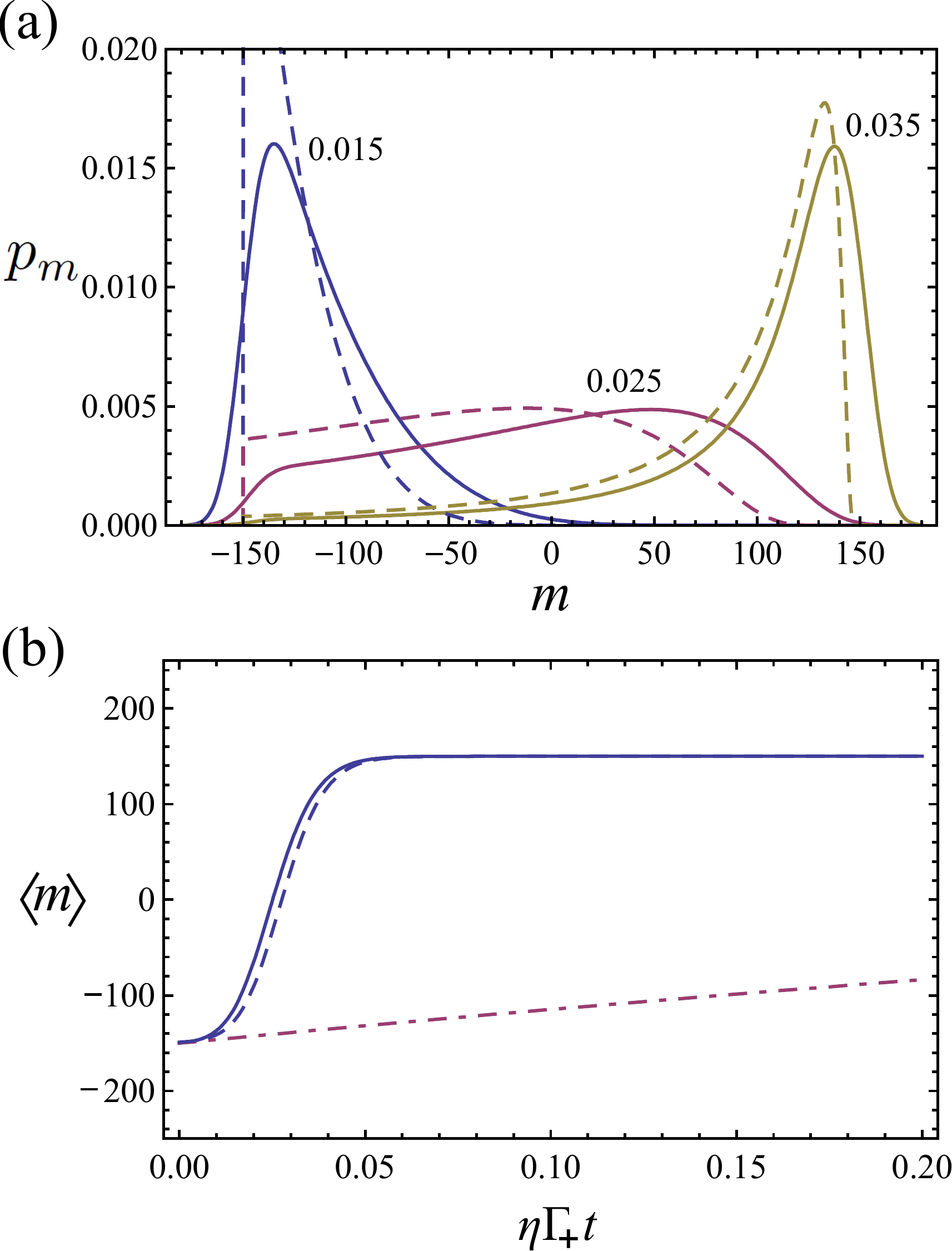}
\caption{\label{superrad_approx_comparison} Panel (a): Comparison of the numerical results for the coherent time evolution of $p_m$ (solid) and the superradiant approximation, Eq.~(\ref{rho_supperadiant}) (dashed). The values of $\eta \Gamma_+ t = 0.015,$ 0.025, and 0.035 are indicated for each of the three curves. Panel (b): plot of the average nuclear magnetization as function of $\eta \Gamma_+ t$ for the coherent (solid) and incoherent (dot-dashed) evolution. The dashed curve is the approximate result obtained from the superradiant distribution, Eq.~(\ref{rho_supperadiant}). For both plots we have used $N=500$ and $R=0.25$, as in Fig.~\ref{ferro_coherent_timeevol}.}
\end{figure}

The discussion above shows that Eq.~(\ref{rho_supperadiant}) can be generally used to describe the nuclear-magnetization dynamics, together with Eqs.~(\ref{Gamma_approx_superradiant}) and (\ref{I_approx_superradiant}). In particular, the time to reach zero polarization is $t_0= \ln(3.28 I)/(2\eta\Gamma I)$. This translates to
\begin{equation}\label{t_0_coh}
\eta \Gamma_+ t_0 \simeq \frac{(1+R)\ln [1.6 N(1-R)/(1+R)]}{N(1-R)^2},~~{\rm (coherent)}
\end{equation} 
which is strongly reduced with $N$, being multiplied by a factor $\ln N/N$. The corresponding value of $\eta \Gamma_+ t_0$
 for the incoherent dynamics obtained from Eq.~(\ref{m_time_incoh}),
\begin{equation}\label{t_0_incoh}
\eta \Gamma_+ t_0 \simeq \frac{\ln 2}{1+R },~~{\rm (incoherent)}
\end{equation} 
is independent of $N$ and can thus become much longer than Eq.~(\ref{t_0_coh}). With the parameters used for Fig.~\ref{ferro_coherent_timeevol}, Eq.~(\ref{t_0_coh}) gives $\eta \Gamma_+ t_0 \simeq 0.03$, in good agreement with the numerical evolution, and Eq.~(\ref{t_0_incoh}) gives $\eta \Gamma_+ t_0 \simeq 0.6$. For a more detailed comparison of the two regimes, and of the superradiant approximation, see Fig.~\ref{superrad_approx_comparison}. 

\subsection{Current dynamics} \label{sec_current}
The interesting behavior of the magnetization dynamics is accessible through the electron current $J$. Neglecting the spin-flip contributions to the current, we obtain the lowest-order result due to sequential tunneling:
\begin{equation}\label{J0}
J_0 = \frac{\Gamma_{+}^{(l)}\Gamma_{-}^{(r)}}{\Gamma_{+}^{(l)}+\Gamma_{-}^{(r)}}.
\end{equation}
Equation \eqref{J0} gives the leading contribution to the constant background current through the spin valve, independent of the nuclear magnetization. On the other hand, the correction $\delta J = J-J_0$ depends on the magnetization dynamics. This distinction is especially clear for half-metal leads, when $J_0=0$ and each tunneling event through the spin valve is associated with a nuclear-spin flip, giving $\delta J = d\langle m \rangle /dt$.

For imperfectly-polarized ferromagnetic leads, the total current $J$ is given by a formula similar to Eq.~(\ref{J0}), but $\Gamma_+^{(l)}$ and $\Gamma_-^{(r)}$ are replaced by the total rates $\Gamma_+^{(l)}+ \Gamma_{\downarrow \to d}^{(l)}[\rho_N]$ and $\Gamma_-^{(r)}+ \Gamma_{d \to \downarrow}^{(r)}[\rho_N]$.  Such an expression for $J$ gives the following lowest-order correction to Eq.~(\ref{J0}):
\begin{align}
\delta J &\simeq (1-n_d)^2 \Gamma^{(l)}_{\downarrow \to d}[\rho_N] +n_d^2 \Gamma^{(r)}_{d \to\downarrow}[\rho_N] \label{dJ_derivation1}\\
& = (1-n_d) \langle \gamma_m^-\rangle + n_d \langle \gamma_m^+ \rangle, \label{dJ_derivation2}
\end{align}
where in the second line we used Eqs.~(\ref{gamma_plus_def}) and (\ref{gamma_minus_def}) (analogous expressions with $\langle \gamma_m^\pm \rangle \to \langle \gamma_{I,m}^\pm \rangle$ apply to the coherent case). $n_d$ is given by Eq.~(\ref{n0_I0_ferrom}), i.e., neglects spin-flip corrections to the dot occupation, which is appropriate for this lowest-order expression for $\delta J$ (in the spin-flip rates). To make a more direct connection to the half-metal limit (when $\delta J = d\langle m \rangle /dt$), we can use $d\langle m \rangle / dt = \langle \gamma_m^+ \rangle - \langle \gamma_m^- \rangle$ and rewrite Eq.~(\ref{dJ_derivation2}) as:
\begin{equation}\label{dJ}
\delta J \simeq n_d \frac{d \langle m \rangle}{dt} + \langle \gamma_m^- \rangle.
\end{equation}
Imperfectly-polarized ferromagnetic leads introduce the multiplicative factor $n_d < 1$ (instead of $n_d=1$). They also result in a finite depolarization rate $\langle \gamma_m^- \rangle$ of the nuclear-spin bath, due to spin-flip tunneling from the left contact (which is absent for half-metal leads). 

For the incoherent evolution, we can compute an explicit result from Eq.~(\ref{dJ}) by using Eq.~(\ref{m_time_incoh}) for $\langle m \rangle$ and Eq.~(\ref{gamma_minus_def}) for $\gamma_m^-$. We obtain:
\begin{align}
\delta J & \simeq \left(n_d -\frac{R}{1+R} \right) \frac{d \langle m \rangle}{dt} + \frac{\eta N \Gamma_-}{1+R}, ~~{\rm (incoherent)} \nonumber \\
&{\rm with} \quad \frac{d \langle m \rangle}{dt} = \eta N \Gamma_+ (1-R) e^{-\eta \Gamma_+ (1+R)t}, \label{dJ_incoh} 
\end{align}
where the second term (the background contribution to $\delta J$) originates from $\langle \gamma_m^- \rangle$. We also notice that the prefactor of $d\langle m \rangle / dt$ is different from $n_d$, due a transient contribution from $ \langle \gamma_m^- \rangle$. Since $d\langle m \rangle / dt$ has a simple exponential decay, the sign of $n_d -R/(1+R)$ determines if the transient current is a decreasing or increasing function of time. Both scenarios are possible: since $n_d=(1+|t_r/t_l|^2\sqrt{R})^{-1}$, the negative sign is realized if $|t_r/t_l|>R^{-3/4}>1$. 

\begin{figure}
\includegraphics[width=0.45\textwidth]{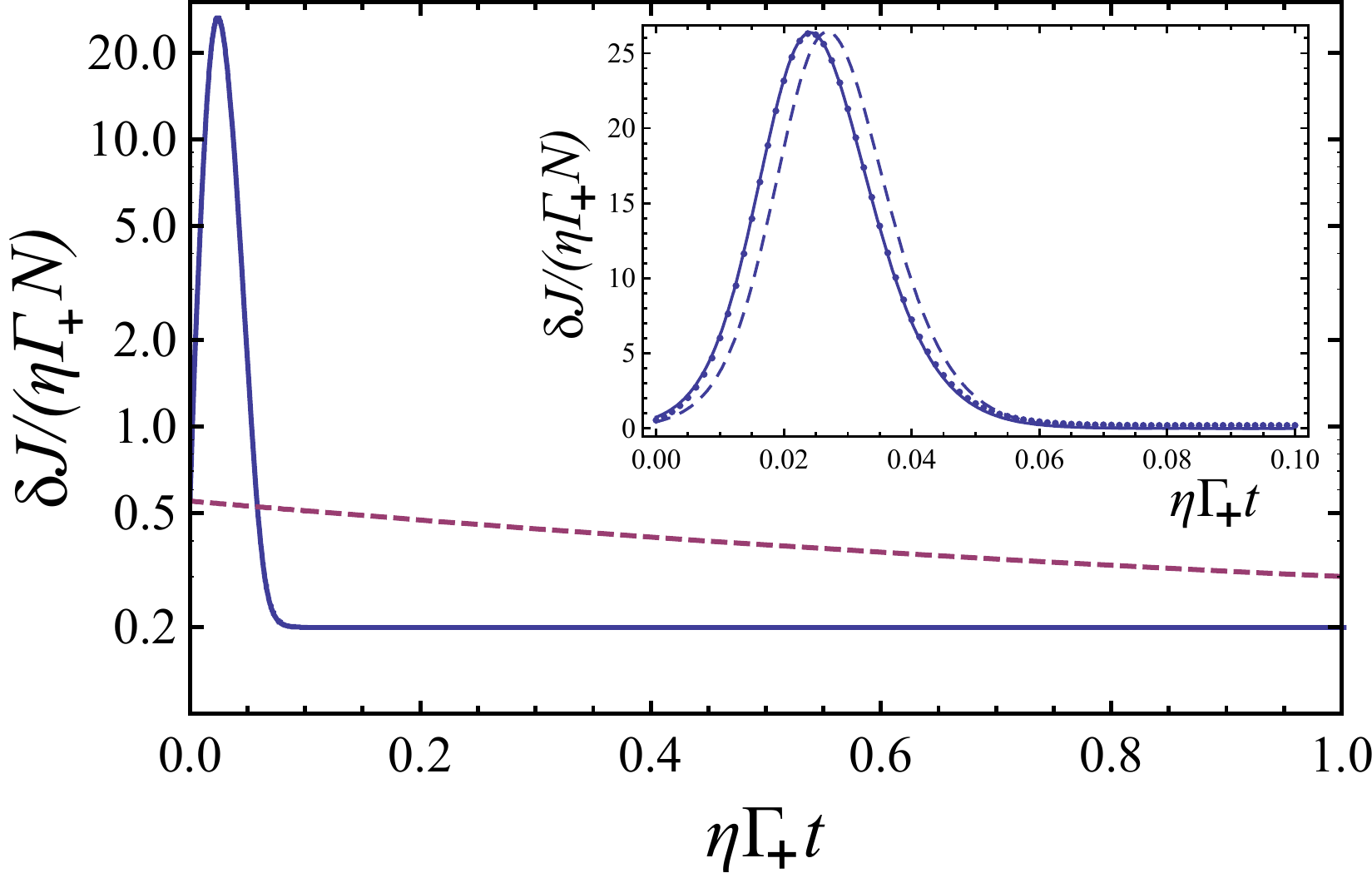}
\caption{\label{current_peak}
The main panel shows a comparison between the coherent (solid) and incoherent (dashed) dynamics of $\delta J$ (notice the logarithmic scale of the $\delta J$ axis). The two curves have the same initial and asymptotic values ($\delta J \to \eta N \Gamma_-/(1+R)$ at large times). The inset shows a comparison of the coherent result, Eq.~(\ref{dJ}) (dots), to the approximate expression, Eq.~(\ref{dJ_coherent}). The solid line is obtained from Eq.~(\ref{dJ_coherent}) using the exact magnetization, while for the dashed line the magnetization is obtained from Eq.~(\ref{rho_supperadiant}) [see also the solid and dashed curve in Fig.~\ref{superrad_approx_comparison}(b)]. In this plot $N=500$, $R=0.25$, and $|t_l/t_r |=1$.}
\end{figure}

The slow monotonic dependence of the current in the incoherent case should be contrasted with the transient current peak of the coherent case, shown in Fig.~\ref{current_peak}. We have discussed in Sec.~\ref{sec_stationary} that the stationary nuclear state $\rho_N$ is the same for the two cases (coherent/incoherent). From Eq.~(\ref{dJ_derivation1}), one concludes that the asymptotic value of the current is the same as that given in Eq.~(\ref{dJ_incoh}). Additionally, at $t=0$ the current is the same in the two cases (see Fig.~\ref{current_peak}) but at intermediate times the time evolution is dramatically different. The coherent case can be approximately described with:
\begin{equation}\label{dJ_coherent}
\delta J \simeq \left(n_d +\frac{R}{1-R}\right) \frac{d \langle m \rangle}{dt} . ~~{(\rm coherent)}
\end{equation}
This formula is obtained from Eq.~(\ref{dJ}) using $d \langle m \rangle/dt = \langle \gamma_{I,m}^+ - \gamma_{I,m}^- \rangle $ and $\gamma_{I,m}^- \simeq R \gamma_{I,m}^+$ (this is justified when the rates with $ I \gg |m|$ are most relevant and we can approximate Eq.~(\ref{gamma_coh_def}) with $\gamma_{I,m}^\pm \simeq \eta \Gamma_\pm (I^2 -m^2)$). As seen in the inset of Fig.~\ref{current_peak}, Eq.~(\ref{dJ_coherent}) is in excellent agreement with the numerical evaluation. A further approximation can be realized by using the superradiant distribution Eq.~(\ref{rho_supperadiant}) to evaluate $d\langle m \rangle/dt$. This gives the dashed curve in the inset of Fig.~\ref{current_peak} and allows us to find analytic expressions for $\delta J$. In particular, the maximum value of the current can be estimated as follows:
\begin{equation}\label{dJ_coherent_peak}
\delta J_{\rm max} \simeq 0.2 \frac{(1-R)^3}{(1+R)^2} \left(n_d +\frac{R}{1-R}\right) \eta N^2 \Gamma_+, 
\end{equation}
and occurs at the same time $t_0$ given by Eq.~(\ref{t_0_coh}). Thus, as in the case of the magnetization dynamics, the time scale of the coherent current pulse is much shorter than for the incoherent dynamics. The current peak has an enhancement factor of order $N$ relative to the incoherent case. This factor is clear comparing Eq.~(\ref{dJ_coherent_peak}) with the typical incoherent value $\sim \eta N \Gamma_+$. These features are analogous to the superradiant emission of light, which is also in the form of a short and intense pulse. At short times, while the incoherent current is essentially constant, the coherent evolution shows an exponential increase $\delta J \propto e^{2 \Gamma I t}$, as also typical for superradiant light emission.\cite{Gross1982, Eto2004}
 
\section{Intermediate nuclear-spin dephasing} \label{sec_mastereq}
In the previous Sections, we have treated two extreme limits: the coherent and incoherent regime in which the nuclear-spin dephasing due to the inhomogeneous term $H_N$ in Eq.~(\ref{Htot}) is either absent or very strong, respectively. We wish now to consider a small but finite degree of dephasing, to establish the robustness of the superradiant-like behavior: the analysis will thus characterize the relevant timescale for nuclear-spin dephasing, below which the superradiant-like transport can be observed.  

\subsection{Crossover timescale}

The question of the relevant time for superradiant-like transport is important since the enhancement of tunneling rates depends on the coherence properties of the nuclear-spin system and the nuclear bath is comprised of a large population of spins, up to $N \sim 10^5 - 10^6$. As is well known, entangled states with a large number of particles have short coherence times, which can scale like $1/N$ (or even $1/N^2$ in the case of spatially correlated phase noise\cite{Monz2012}). However, we find no unfavorable scaling with $N$ in this case.

To understand the final result in simple terms, we refer to the expressions for the spin-flip tunneling rates, Eqs.~(\ref{tunnel_rate_flipL1}) and (\ref{tunnel_rate_flipR1}). For an angular-momentum eigenstate $|N/2, m\rangle$, all the $\langle n_m |\rho_N|n'_m\rangle$ have the same value but are affected very differently by the inhomogeneous broadening $\Delta b$: high-order coherences $\langle n_m |\rho_N|n'_m\rangle$, where $|n_m\rangle$, $|n'_m\rangle$ differ on a large number of nuclear spins, typically decay with a large rate $\sim N\Delta b$. However, Eq.~(\ref{tunnel_rate_flipR1}) depends on $\rho_N$ through the factor ${\rm Tr}[I_+ I_- \rho_N]$, which only involves the \emph{second-order} coherences, i.e., where $|n_m\rangle$, $|n'_m\rangle$ differ by a single flip-flop process. This suggests that the relevant time scale to observe the superradiant-like transport is given by
\begin{equation}\label{t_phi_scale}
\tau_\phi \sim (\Delta b)^{-1}.
\end{equation}
This argument indicates that $\tau_\phi$ does not scale with $N$ and is of the same order as the dephasing time for a single nuclear spin (nuclear-spin coherence times of $\sim 1$~ms have been reported in quantum dots,\cite{Takahashi2011} and up to $\sim 3$ hours for ionized donors in silicon\cite{Saeedi2013}), thus that nuclear-spin decoherence does not pose a severe limitation to observing the `superradiant' transport regime with large $N$.  We will confirm this result below with a more sophisticated calculation.

\subsection{Nuclear-spin master equation}

To derive a master equation for the nuclear-spin dynamics, we find it simpler to start from a transformed Hamiltonian, obtained by applying:
\begin{equation}
U = e^{\frac{A}{2N b}(S_+I_- -S_- I_+)},
\end{equation}
which eliminates the flip-flop terms of the hyperfine coupling $H_{ff}$ to lowest order in $A/b$. We obtain:
\begin{equation}\label{H_rotated}
U H U^\dag  \simeq (H_{\rm el} + H_T + H_N) +H_I,
\end{equation}
where $H_I=H_{zz}^{\rm hf}+\delta H_T$ is the interaction Hamiltonian. $\delta H_T$ describes the following flip-flop tunneling processes:
\begin{equation}\label{dH_T_def}
\delta H_T = - \frac{A}{2N b} \sum_p ( t_l l_{p\downarrow}^\dag + t_r r_{p\downarrow}^\dag ) d_\uparrow I_+ + {\rm h.c.} ,
\end{equation}
arising from the original tunneling Hamiltonian, $U H_T U^\dag \simeq H_T +\delta H_T$. In $H_I$, in addition to keeping only contributions up to first order in $A/b, b_k$ (we suppose $b_k \sim A/N$), we have also omitted terms involving $d^{(\dag)}_\downarrow$ in $\delta H_T$. These terms have a negligible influence in the regime considered here since $d^{(\dag)}_\downarrow|0\rangle$ is outside the (large) bias window. Using Eq.~(\ref{dH_T_def}), it is sufficient to consider the lowest-order Fermi's golden rule to obtain spin-flip tunneling rates in agreement with Eqs.~(\ref{tunnel_rate_flipL1}) and (\ref{tunnel_rate_flipR1}).

We now consider a natural partition of the Hamiltonian, Eq.~(\ref{H_rotated}), into electronic and nuclear-spin degrees of freedom, with $H_0 = H_{\rm el}+ H_T+ H_N$ and the interaction Hamiltonian $H_I$. Since the electron dynamics are generally much faster than the nuclear-spin dynamics, we derive an approximate master equation for the nuclear-spin bath starting from the standard Born-Markov approximation:\cite{Blum2012,Breuer2007} 
\begin{align}\label{master_blum}
\dot{\tilde \rho}_{N}(t) = & -i \, {\rm Tr}_{\rm el} [\tilde{H}_I(t),{\tilde \rho}_{N}(0) \otimes \rho_{\rm el}] \nonumber\\
& - {\rm Tr}_{\rm el} \int_0^\infty d\tau [\tilde{H}_I(t),[\tilde{H}_I(t-\tau),{\tilde \rho}_{N}(t)\otimes \rho_{\rm el}]],
\end{align}
where ${\rm Tr}_{\rm el}[\ldots ]$ indicates a trace with respect to the electronic degrees of freedom and $\tilde O(t) $ are operators in the interaction picture. In particular:
\begin{equation}\label{I_tilde_pm}
\tilde{I}_\pm (t) = \sum_k I_{k,\pm} e^{\pm i b_k t}.
\end{equation}
Equation~(\ref{master_blum}) assumes a factorized density matrix $\tilde\rho(t) \simeq \tilde\rho_N(t) \otimes \rho_{\rm el}$, where $\rho_{\rm el}$ is the stationary electron state. Taking into account the tunneling process ($H_T$ is included in $H_0$): 
\begin{equation}\label{rho_el}
\rho_{\rm el} = n_d d_\uparrow^\dag | 0 \rangle \langle 0 | d_\uparrow +(1-n_d)| 0 \rangle \langle 0 |,
\end{equation}
where $n_d$ is given by Eq.~(\ref{n0_I0_ferrom}). Working in a weak-tunneling regime, we have neglected the effect of tunneling on the electronic states $|0\rangle,d_\uparrow^\dag | 0 \rangle $ appearing in Eq.~(\ref{rho_el}) and, within the same range of validity, we neglect $H_T$ in the interaction picture operators. As a result, $\tilde{l}_{p\downarrow}(t)=l_{p\downarrow} e^{-i \epsilon^{(l)}_{p\downarrow} t}$, $\tilde{r}_{p\downarrow}(t)=r_{p\downarrow}e^{-i \epsilon^{(r)}_{p\downarrow} t}$, which allow one to evaluate Eq.~(\ref{rho_el}) in a straightforward way \cite{Blum2012,Breuer2007}. We also set $\tilde d_\uparrow (t)= d_\uparrow $, by choosing $V_g=b/2$ in Eq.~(\ref{H_ideal}).  

A further simplification in the final form is achieved with ${\tilde \rho}_{N}(t)$ diagonal in the $I_z$ quantum number. This requirement is physically justified since it is consistent with the final form of the nuclear-spin master equation and, as discussed in Sec.~\ref{sec_stationary}, we can safely assume that in the initial state any coherence in $I_z$ has decayed to zero. Under these assumptions, and
returning to the Schr\"odinger picture:
\begin{equation}\label{master_equation}
\dot\rho_{N}(t)= -i [H_N+H_{LS}, \rho_N(t)] + \sum_\pm \eta \Gamma_\pm \, \mathcal{D}[I_\pm]\rho_{N}(t),
\end{equation}
with the standard Lindblad dissipator $\mathcal{D}[A] \rho = A\rho A^\dag -\frac12 (A^\dag A \rho + \rho A^\dag A) $\cite{Breuer2007} and the Lamb-shift Hamiltonian:
\begin{equation}\label{LS}
H_{LS} = \frac12 \eta \Gamma_{LS} I^2,
\end{equation}
where, assuming a uniform density of states within a symmetric bandwidth $|\epsilon^{(\alpha)}_{p\downarrow}|<\Delta_\alpha$:
\begin{equation}\label{LS_frequency}
\Gamma_{LS} = \frac{\Gamma_-^{(l)}}{\pi}\ln\left|\frac{\mu_l}{\Delta_l} \right|+\frac{\Gamma_+^{(r)}}{\pi}\ln\left|\frac{\mu_r}{\Delta_r} \right|.
\end{equation}

The final form of the unitary dynamics in Eq.~(\ref{master_equation}) can be understood by noticing that, working with a $\rho_N$ diagonal with respect to $I_z$, the contribution from $H_{zz}^{\rm hf}$ is absent. The contribution of $\delta \tilde{H}_T(t)$ is also zero, due to the average over the electronic state. As for the dissipator, since Eq.~(\ref{dH_T_def}) does not conserve the occupation number of the quantum dot, all terms in Eq.~(\ref{master_blum}) that are linear in $\delta \tilde{H}_T(t)$ vanish. This implies that the two perturbations $\delta {H}_T$ and $H_{zz}^{\rm hf}$ act independently on the time evolution, with no mixed term appearing on the right-hand side of Eq.~(\ref{master_blum}). If $[I_z, {\tilde \rho}_{N}(t)]=0$, the only nonvanishing contribution from $\tilde{H}_I(t)$ is due to  $\delta \tilde{H}_T(t)$, which yields the final result given in Eq.~(\ref{master_equation}). As seen, Eq.~(\ref{master_equation}) is consistent with our assumption $[I_z, {\tilde \rho}_{N}(t)]=0$ since both $H_N$ and $H_{LS}$ commute with $I_z$ and the dissipators $\mathcal{D}[I_\pm]$ preserve the diagonal form of $\rho_N$ as well.

As a final remark on Eq.~(\ref{master_equation}), we discuss the Lamb-shift term Eq.~(\ref{LS}). In the derivation we have neglected small non-uniform terms and used again the fact that $I_\pm I_\mp$ can be replaced by $I^2$, if $[I_z, {\tilde \rho}_{N}(t)]=0$.  A few more details are given in Appendix \ref{sec:appendix}. The logarithmic divergence in Eq.~(\ref{master_equation}) at large bandwidth would be cut-off by treating the time evolution beyond the Born-Markov approximation of Eq.~(\ref{LS_frequency}). However, we refrain ourselves from a full microscopic derivation and consider $\Gamma_{LS}$ as a phenomenological parameter. This is also justified because terms similar to $H_{LS}$ would appear by including higher orders in $A/b$ in the rotated Hamiltonian, Eq.~(\ref{H_rotated}).

\subsection{Numerical results}

We now solve Eq.~(\ref{master_equation}) numerically, for values of $b_k$ chosen from a Gaussian distribution with standard deviation $\Delta b$. By focusing on the half-metal regime ($\Gamma_-=0$), an additional simplification arises in the numerical solution: since the equations for $\langle n_m| \rho_N | n'_m \rangle$ only depend on $\langle n_{m-1}| \rho_N | n'_{m-1} \rangle$, the problem can be solved iteratively for $m+N/2=0,1,2,\ldots$. We assume a fully polarized nuclear state at $t=0$, which gives $p_{-N/2}(t)=e^{-\eta \Gamma_+ N t}$. 

\begin{figure}
\includegraphics[width=0.45\textwidth]{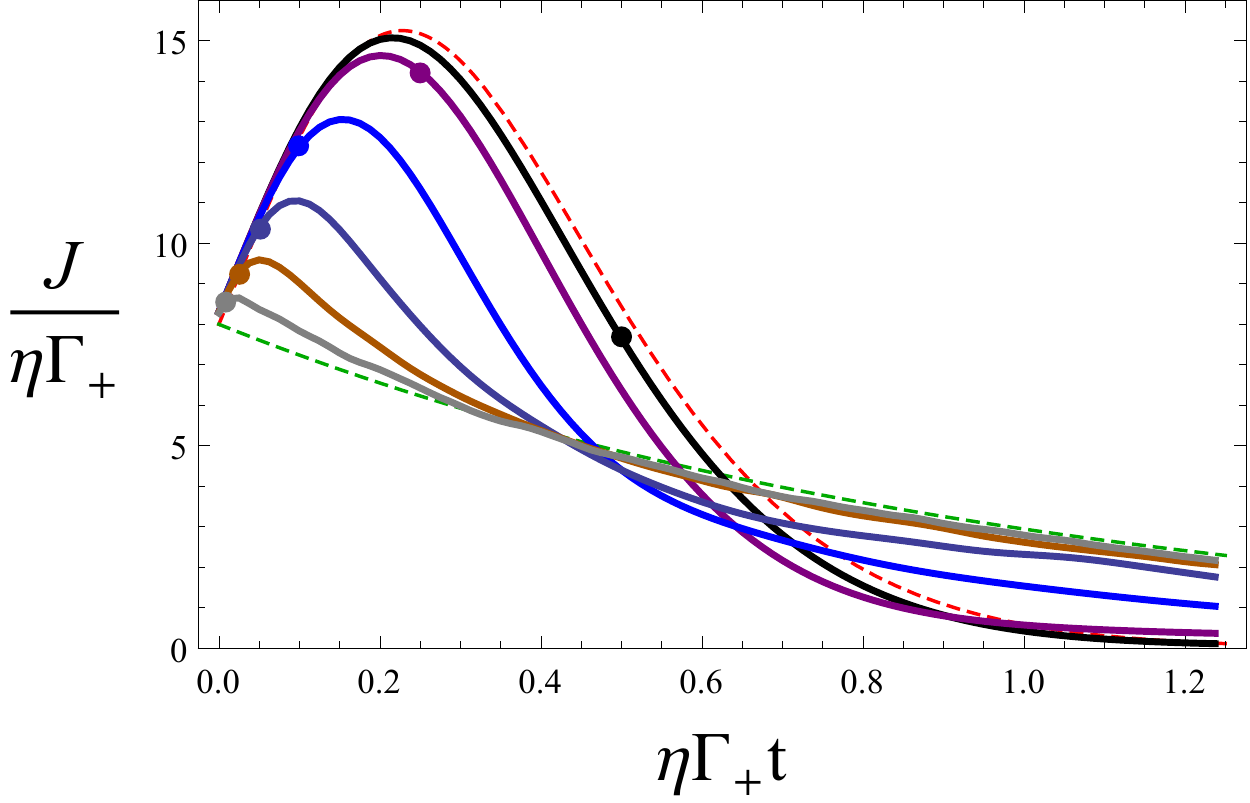}
\caption{\label{current} Electron current through a quantum dot in contact with half-metal leads and 8 nuclear spins. The upper red dashed curve is for $\Delta b =0$. Solid curves with $\Delta b/(\eta \Gamma_+) = 1,2,5,10,20,50$ progressively approach the $\Delta b \to \infty$ result (lower green dashed curve). The dots mark $t=\frac12 \Delta b^{-1}$ for each of the $\Delta b$ values. $\Gamma_{LS}=0$ and each solid curve is an average over 20 realizations of the local nuclear fields. }
\end{figure}

Figure~\ref{current} shows an example of the current through the quantum dot for different values of $\Delta b$ (due to the half-metal leads, the electron current can be obtained as $J=d\langle I_z \rangle/dt$). As expected from Eq.~(\ref{t_phi_scale}), a small/large value of $\Delta b$ allows to recover the coherent/incoherent solutions discussed in detail in the previous sections. In particular, it is shown that below a timescale $t=\frac12 \Delta b^{-1}$ (dots), appropriate for the decay of the second-order coherence, all curves are well approximated by the coherent evolution. In Figure~\ref{current}, the deviation of the current from  the incoherent limit (a simple exponential decay) can be easily identified, which demonstrates how the effect of nuclear-spin coherence is already evident in small nuclear-spin clusters (in this example, 8 nuclear spins). 

\begin{figure}
\includegraphics[width=0.45\textwidth]{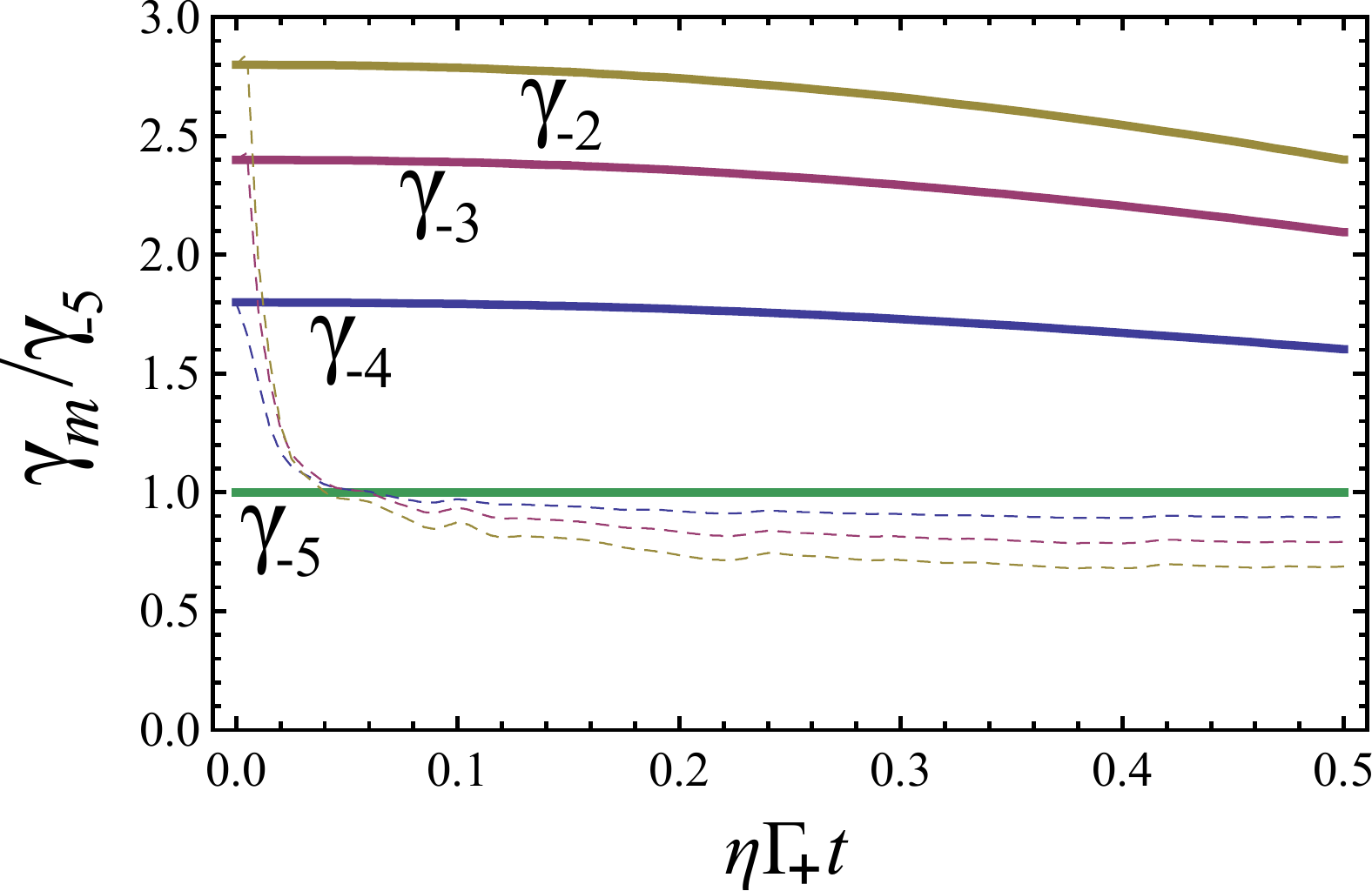}
\caption{\label{fig_rates} Instantaneous transition rates $\gamma_m(t)$ for a system with half-metal leads and $N=10$ nuclear spins (initial nuclear state is $|5,-5\rangle$) and $\Gamma_{LS}=0$. Solid lines are for $\eta\Gamma_+/\Delta b = 0.5$ and show a relatively small decay from the $t=0$ coherent rates $\eta \Gamma_+(5-m)(6+m)$. The dashed lines are for  $\eta\Gamma_+/\Delta b = 5\times 10^{-3}$. On the $\tau_\phi\sim \Delta b^{-1}$ timescale, they show full decay to the incoherent results $\eta \Gamma_+(5-m)$.}
\end{figure}

Of course, the master equation yields much more information on the statistical properties of the system dynamics than the total current. We focus now on the tunneling rates to confirm the behavior obtained perturbatively in Eqs.~(\ref{tunnel_rate_flipL1}) and (\ref{tunnel_rate_flipR1}). If individual tunneling events are detected, the observer can keep track of the nuclear spin polarization at time $t$ in each single run (with half-metal leads and starting from $|N/2,-N/2\rangle$). Under this scenario, we introduce the time-dependent tunnel rates $\gamma_m(t)$, i.e., the tunnel rates from source to drain, conditional on having nuclear polarization $m$ at time $t$. The $\gamma_m(t)$ depend on the instantaneous nuclear state $\rho_N(t)$ and, based on Eqs.~(\ref{tunnel_rate_flipL1}) and (\ref{tunnel_rate_flipR1}), it is natural to expect the following behavior:
\begin{equation}\label{gamma_mastereq_limits}
\gamma_m(t)=\left\{
\begin{array}{ll}
\gamma^+_{N/2, m} & {\rm for}~ t\ll \tau_\phi, \\
\gamma^+_m & {\rm for}~ t\gg \tau_\phi,
\end{array}
\right.
\end{equation}
where $\gamma^+_{N/2, m},\gamma^+_m$ are defined in Eqs.~(\ref{gamma_coh_def}) and (\ref{gamma_plus_def}), respectively. In other words, at short/long times, the instantaneous tunnel rates yield the coherent/incoherent results of earlier sections.  We can obtain the $\gamma_m(t)$ by solving the detailed balance relation:
\begin{equation}\label{gamma_mastereq}
\dot{p}_m(t)=\gamma_{m-1}(t)p_{m-1}(t)-\gamma_{m}(t)p_{m}(t),
\end{equation}
where $p_m(t)$ are obtained from Eq.~(\ref{master_equation}). As seen in Fig.~\ref{fig_rates}, the $\gamma_m(t)$ satisfy Eq.~(\ref{gamma_mastereq_limits}). At intermediate times, partial dephasing of $\rho_N$ entering Eqs.~(\ref{tunnel_rate_flipL1}) and (\ref{tunnel_rate_flipR1}) interpolates between the two results. The numerical evaluation of $\gamma_m (t)$ in Fig.~\ref{fig_rates} shows quite clearly the crossover between the two regimes, and that all the $\gamma_{m}(t)$ approach the incoherent result around $t\sim \tau_\phi \sim \Delta b^{-1}$.

\begin{figure}
\includegraphics[width=0.45\textwidth]{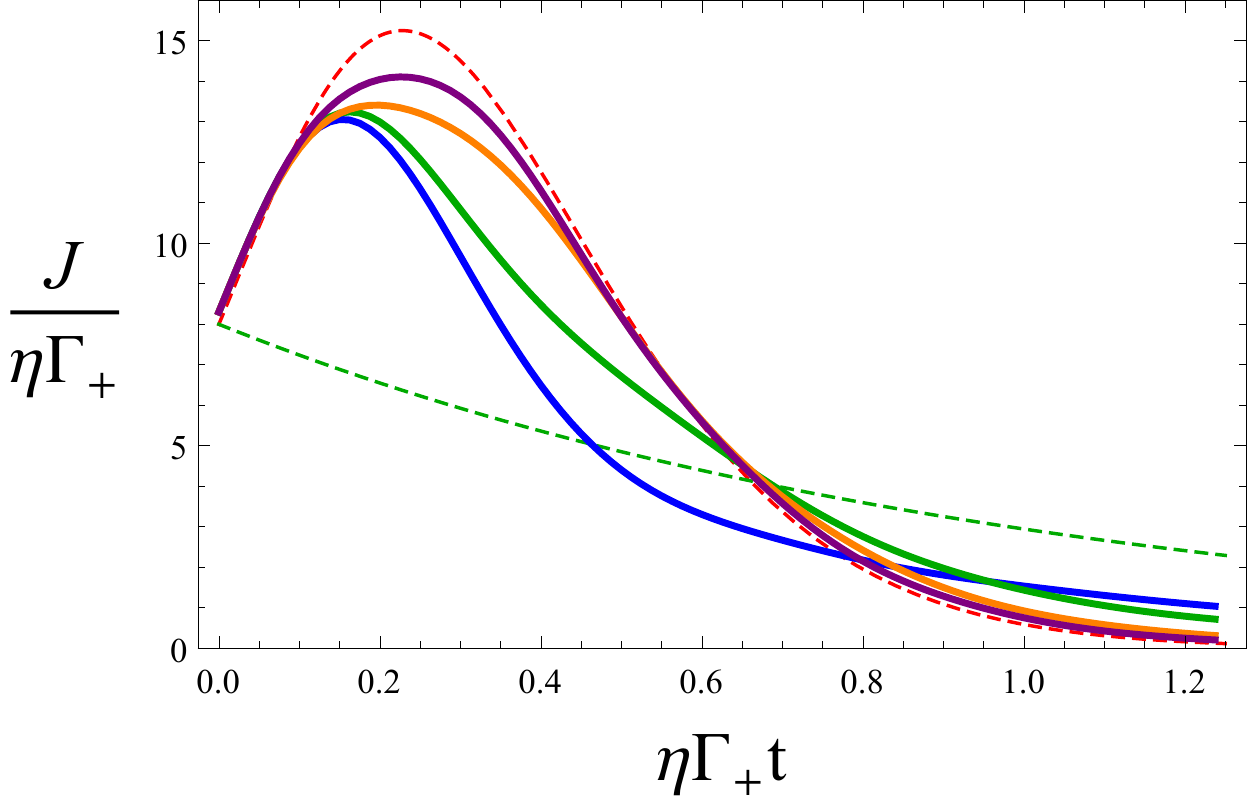}
\caption{\label{lamb_shift} Effect of the Lamb shift, Eq.~(\ref{LS}), for a system of $N=8$ nuclear spins in contact with half-metal leads and $\Delta b/(\eta\Gamma_+) =5$. The solid curves are with $\Gamma_{LS}/(\eta\Gamma_+) =0,2,4,5$, approaching the $\Delta b =0$ result (upper dashed curve) for larger values of $\Gamma_{LS}/(\eta\Gamma_+)$. The lower dashed curve is the incoherent result ($\Delta b = \infty$)  }
\end{figure}

While in Figs.~\ref{current} and \ref{fig_rates} we assumed for simplicity $\Gamma_{LS}=0$, we consider now the effect of a finite Lamb-shift. It was already discussed that terms $\propto I^2$, see Eq.~(\ref{LS}), make the coherent evolution more robust.\cite{Schuetz2012} This behavior is demonstrated in Fig.~(\ref{lamb_shift}) where larger values of $\Gamma_{LS}$ modify the current dynamics and enhance the current peak, which approaches the coherent superradiant-like result. 

\section{Conclusion} \label{sec_concl}

We have discussed transport through a quantum dot in contact with ferromagnetic leads, for a spin valve configuration realized by several recent experiments.\cite{Hamaya2007,Hamaya2008a,Hamaya2008b,Tarun2011,Dirnaichner2015} Focusing on the hyperfine-mediated flip-flop processes, we have analyzed two distinct regimes of coherent/incoherent evolution. In the coherent limit the nuclear-spin system is quickly driven into dark states \cite{Imamoglu2003} and only a small amount of polarization can be generated, starting from an unpolarized thermal state. However, nuclear-spin dephasing can drive the nuclear-spin system away from such dark states, resulting in a sizable nuclear polarization in the long-time limit. This stationary value of the nuclear polarization is simply given by the polarization of the leads. By inverting the bias of the spin valve, the sign of nuclear polarization can be inverted through a fast coherent dynamics. The transport current in this coherent regime reveals features analogous to the superradiant light emission,\cite{Gross1982} as considered theoretically in several recent works.\cite{Eto2002,Eto2004,Schuetz2012,Lai2015} In particular, the enhancement of the transient current by a large factor of order $\sim N$ (the number of nuclear spins) is due to the creation of long-range coherence in the nuclear-spin system, which thus can be directly monitored through electron transport.

We also analyzed the crossover between the two (coherent/incohernt) regimes, showing that the collective enhancement survives on a relatively long timescale, i.e., the dephasing time of individual nuclear spins. Furthermore, it is well known that the largest collective enhancement occurs in the limit of uniform coupling constants. We have outlined a realistic strategy to realize this limit in nanowire quantum dots, through the fabrication of a ``nuclear-spin island'' of spinful isotopes, complemented by wavefunction engineering in the longitudinal and radial directions. The latter can be realized through a suitably tailored core-shell structure. We expect that the implementation of such ``box model'' hyperfine interaction, by allowing a controlled interaction of the electron spin to the nuclear-spin system, could find many interesting applications well beyond the transport model studied here.

\begin{acknowledgments}
We thank J.~Baugh, F.~Kuemmeth, O.~Moutanabbir, and Y.~Tokura for useful discussions. We acknowledge financial support from NSERC, CIFAR, INTRIQ, and FRQNT.
\end{acknowledgments}

\appendix

\section{Exact hyperfine eigenstates with uniform coupling}\label{sec:exact_eigenstates}

We summarize here the exact solution of the quantum dot Hamiltonian with uniform hyperfine coupling, $ H_{\rm hf}+H_{\rm el}$, see Eqs.~(\ref{hf}) and (\ref{H_ideal}). If $|0\rangle$ describes the electronic state with an empty dot, the eigenstates are simply given by $|0\rangle|I, m \rangle$ (suppressing an additional permutation index, which does not enter the eigenvalues\cite{Arecchi1972}), where $I(I+1)$ and $m$ are the eigenvalues of $\mathbf{I}^2$ and $I_z$, respectively. For a singly-occupied dot we have the following eigenstates (with $m=-I\ldots I-1$):
\begin{align}\label{exact_eigenstates1}
\left|\varphi_I^+(m)\right> &=\alpha_{I,m}d^\dag_\uparrow|0\rangle |I, m \rangle+\beta_{I,m}d^\dag_\downarrow|0\rangle |I, m+1 \rangle,\\
\left|\varphi_I^-(m)\right> &=\alpha_{I,m}d^\dag_\downarrow|0\rangle |I, m +1 \rangle-\beta_{I,m}d^\dag_\uparrow|0\rangle |I, m \rangle,\label{exact_eigenstates2}
\end{align}
where $\alpha_{I,m}=\cos(\theta_{I,m}/2)$ and $\beta_{I,m}=\sin(\theta_{I,m}/2)$, with
\begin{eqnarray}
\label{tan_theta}
\theta_{I,m}= \arg\bigg[ &&-\frac{b}{2}+\frac{A}{4N}(2m+1)  \nonumber \\
   && + i \frac{A}{2N} \sqrt{I(I+1)-m(m+1)}\bigg].
\end{eqnarray}
The corresponding energies are:
\begin{equation}\label{energies_dot_exact}
\epsilon_{I,m}^\pm =V_g-\frac{A}{4N}\pm\sqrt{\frac{b^2}{4}-\frac{bA}{4N}(2m+1)+\left[\frac{A}{4N}(2I+1)\right]^2}.
\end{equation}
The states in Eqs. \eqref{exact_eigenstates1} and \eqref{exact_eigenstates2} are supplemented by the fully-polarized eigenstates $d^\dag_{\uparrow}|0\rangle |I, I \rangle$, $d^\dag_{\downarrow}|0\rangle |I, - I \rangle$, with eigenvalues $V_g \mp  b/2+ A I/(2N)$.

An interesting limiting result is when the external magnetic field is zero. By setting $b=0$ and assuming for definiteness $A>0$, Eqs.~(\ref{tan_theta}) and (\ref{energies_dot_exact})   yield
\begin{align}\label{coefficients_small_b}
&\alpha_{I,m}=\sqrt{\frac{I+m+1}{2I+1}},\quad \beta_{I,m}=\sqrt{\frac{I-m}{2I+1}},\\
&\epsilon_{I,m}^\pm =V_g-\frac{A}{4N}\pm\frac{A}{2N}(I+1/2),\quad {\rm at~}b=0, \label{energies_large_b}
\end{align}

In the opposite regime of large magnetic field:
\begin{align}\label{coefficients_large_b}
&\alpha_{I,m}\simeq \frac{A}{2Nb}\sqrt{I(I+1)-m(m+1)},\quad \beta_{I,m}\simeq 1,\\
&\epsilon_{I,m}^\pm \simeq V_g-\frac{A}{4N}\pm \left[\frac{b}{2}- \frac{A}{2N}(m+1/2)\right],\quad {\rm at~}b\gg A, 
\end{align}
where in the above expressions we have assumed $b>0$ and neglected terms of higher order in $A/b$.

\section{Lamb shift Hamiltonian}\label{sec:appendix}

By explicitly writing the $\tau$ dependence of the integrand, the second line of Eq.~(\ref{master_blum}) gives:
\begin{eqnarray}\label{master_equation_1}
\sum_{\alpha,p,k,k'} \eta \int_0^\infty d\tau |t_\alpha|^2 \Big\{ n_d (1-n_{\alpha,p}) \mathcal{D}_{k,k'}^{(+)}[\tilde{\rho}_{N}(t)] \nonumber \\
+  (1-n_d)n_{\alpha,p} \mathcal{D}_{k,k'}^{(-)}[\tilde{\rho}_{N}(t)] \Big\} e^{i (\epsilon^{(\alpha)}_{p\downarrow}-b_{k'})\tau} +{\rm h.c.}, \quad
\end{eqnarray}
where $\alpha \in \{l,r \}$ labels the two leads. In Eq.~(\ref{master_equation_1}) we have introduced the lead occupation numbers $n_{\alpha,p} = \theta(\mu_\alpha-\epsilon^{(\alpha)}_{p\downarrow})$ (the applied bias is $\Delta \mu= \mu_l - \mu_r$, see Fig.~\ref{transport}) and $\mathcal{D}_{k,k'}^{(\pm)}$ are defined as follows (all operators at time $t$):
\begin{align}
\mathcal{D}_{k,k'}^{(+)}[\tilde\rho_{N}(t)] = \tilde{I}_{k',+}\,\tilde\rho_{N}\, \tilde{I}_{k,-} - \tilde{I}_{k,-}\tilde{I}_{k',+}\tilde\rho_{N} ,\\
\mathcal{D}_{k,k'}^{(-)}[\tilde\rho_{N}(t)] = \tilde{I}_{k,-} \,\tilde\rho_{N} \, \tilde{I}_{k',+} -\tilde\rho_{N} \, \tilde{I}_{k',+} \tilde{I}_{k,-}.
\end{align}
The terms of Eq.~(\ref{master_equation_1}) proportional to $n_{\alpha,p}$ can be evaluated as follows (the remaining terms with $1-n_{\alpha,p}$ can be treated in the same way):
\begin{align}\label{useful_integral}
&\sum_p \int_0^\infty n_{\alpha,p} e^{\pm i (\epsilon^{(\alpha)}_{p\downarrow}-b_{k'})\tau} d\tau \nonumber \\
=&\nu_{\alpha\downarrow}\int_{-\Delta_\alpha}^{\Delta_\alpha} d\epsilon^{(\alpha)}_{p\downarrow} n_{\alpha,p} \left[\pi \delta(\epsilon^{(\alpha)}_{p\downarrow}-b_{k'})\pm i {\rm P}\frac{1}{\epsilon^{(\alpha)}_{p\downarrow}-b_{k'}} \right],
\end{align}
where we have written $\sum_p \simeq \nu_{\alpha\downarrow}\int_{-\Delta_\alpha}^{\Delta_\alpha} d\epsilon^{(\alpha)}_{p\downarrow}$, by assuming a symmetric band with constant density of states ($\nu_{l\downarrow}=\nu_-$ and $\nu_{r\downarrow}=\nu_+$). Since $b_k \ll |\mu_{\alpha}|$, the integration of $\delta(\epsilon^{(\alpha)}_{p\downarrow}-b_{k'})$ in the second line is immediate. It gives $\pi \nu_- \delta_{\alpha,l}$, which contributes to the Lindblad dissipator of Eq.~(\ref{master_equation}). 

On the other hand, the integration of the second term in the square parenthesis of Eq.~(\ref{useful_integral}) evaluates to $\pm i \nu_{\alpha \downarrow} \ln\left|\frac{\mu_\alpha-b_{k'}}{\Delta_\alpha+b_{k'}}\right|$ and contributes to the Lamb shift $H_{LS}$ appearing in the first term of Eq.~(\ref{master_equation}). The presence of $b_{k'}$ makes the result inhomogeneous with respect to the nuclear index $k'$. However, notice that the effect of $b_{k'}$ on $H_{LS}$ is small since $\ln |\frac{\mu_\alpha - b_{k'}}{\Delta_\alpha + b_{k'}}| \simeq \ln|\mu_\alpha / \Delta_\alpha|-b_{k'}/\mu_\alpha$ and $b_{k'}/\mu_\alpha \ll 1$. Therefore, the corrections are smaller than $\| H_{LS} b_{k'} /\mu_\alpha\| \ll \|H_{LS}\|,\|H_N \|$, and it is justified to neglect them. It then becomes straightforward to evaluate $H_{LS}$ from Eq~(\ref{master_equation_1}):
\begin{align}\label{LS1}
H_{LS}=& \frac12 \eta \Gamma_{LS}  \frac{\Gamma_-^{(r)} I_+ I_-+ \Gamma_+^{(l)}I_- I_+}{\Gamma_-^{(r)}+\Gamma_+^{(l)}} ,
\end{align}
where $\Gamma_{LS}$ is defined in Eq.~(\ref{LS_frequency}). As discussed in the main text, our assumption on $\rho_N$ allows us to substitute $I_\pm I_\mp \to I^2$ in Eq.~(\ref{LS1}), which leads to the result cited in Eq.~(\ref{LS}).

\bibliography{bibfile}
\end{document}